\documentstyle[12pt]{article}
\begin{document}
\makeatother
\renewcommand{\theequation}{\thesection.\arabic{equation}}
\newcommand{\am}{a \times m}
\newcommand{\ri}{RI }
\newcommand{\mathleftline}[1]{\hbox to0pt{\hss\hbox to12pt{\hbox
 to\hsize{$#1$\hfill}\hss}}}
\newcommand{\COM}[1]{[[[#1]]]}
\newcommand{\ot}{\frac{1}{2}}
\newcommand{\D}{& \displaystyle} 
\newcommand{\di}{\displaystyle} 
\font \math=msbm10 scaled \magstep 0
\newcommand{\mmath}[1]{{\mbox{\math #1}}}
\newcommand{\mycheck}[1]{\marginpar{#1}}
\newcommand{\C}{\cal{C}}
\renewcommand{\Re}{\mathop{\rm Re}}         
\renewcommand{\Im}{\mathop{\rm Im}}         
\newcommand{\half}{{1\over2}}               
\newcommand{\quarter}{{1\over4}}            
\newcommand{\eighth}{{1\over8}}             
\newcommand{\third}{{1\over3}}              
\newcommand{\sixth}{{1\over6}}              
\newcommand{\beq}{\begin{equation}}
\newcommand{\eeq}{\end{equation}} 
\newcommand{\dleft}{\stackrel{\leftarrow}{D}}
\newcommand{\dright}{\stackrel{\rightarrow}{D}}
\newcommand{\heading}[1]{\medskip\noindent{\bf #1}\newline\medskip}
\newcommand{\tra}{\hbox{Tr}}
\newcommand{\trln}{\hbox{Tr}\ln}
\renewcommand{\topfraction}{0.9} 
\date{January 1996}
\title
{The Kosterlitz--Thouless Universality Class}
\author
{\bf R. Kenna\thanks{Supported by EU Human Capital 
and Mobility Scheme Project No. CHBI--CT94--1125}~
and A.C. Irving   \\
~\\ 
Department of Mathematical Sciences,\\ Theoretical Physics Division,\\
University of Liverpool L69 3BX, England}
\maketitle
\begin{abstract}
We examine the Kosterlitz--Thouless universality class and show 
that essential scaling at this type of phase transition is not 
self--consistent unless multiplicative logarithmic corrections 
are included. In the case of specific heat these logarithmic 
corrections are identified analytically. To identify those 
corresponding to the susceptibility we set up a numerical method 
involving the finite--size scaling of Lee--Yang zeroes. We also 
study the density of zeroes and introduce a new concept called 
index scaling. We apply the method to the $XY$--model and the 
closely related step model in two dimensions. 
The critical parameters (including  logarithmic corrections) of 
the step model are compatable with those of the $XY$--model  
indicating  that both models belong to the same universality 
class. This result then raises questions over how a vortex binding 
scenario can be the driving mechanism for the phase transition. 
Furthermore, the logarithmic corrections identified numerically 
by our methods of fitting
are not in agreement with the renormalization group
predictions of Kosterlitz and Thouless.
\end{abstract}

\newpage

\section{Introduction}

\setcounter{equation}{0}

In lattice field theory one is interested in the phenomenon of phase 
transitions, where the quantity representing the length scale of 
relevant physics (correlation length or inverse mass gap) diverges. 
Typically, in spin models, the transition is between phases with 
and without long range order. In models with a temperature driven 
phase transition these phases are distinguished by negative and 
positive reduced temperature ($t$) respectively. A conventional 
phase transition is then characterised by a power--law  divergence 
in the (infinite volume) correlation length near criticality 
(near $t=0$);
\begin{equation}
 \xi_\infty  \sim  \mid t \mid^{-\nu}
\quad .
\label{convxi}
\end{equation}   

\subsection{The $XY$--Model}

In two dimensional spin models with continuous symmetry group and 
continuous interaction Hamiltonian, the existence of a phase with 
conventional long range order is precluded by the Mermin--Wagner 
theorem \cite{MW}. Therefore there is no spontaneous magnetisation 
in $O(n)$--spin models for $n \ge 2$. Physically, the reason for 
this is that any long range order which would otherwise be present 
is destroyed by spin wave excitations in two dimensions.

In such models there can however exist topological long range order.
Indeed, in a $d$-dimensional theory, if the order parameter lies in 
a space $G$, then topological defects of dimension $p$ can occur if 
the $(d-p-1)^{\rm{th}}$ homotopy group $\pi_{d-p-1}(G)$ is non 
trivial \cite{Coleman}. Thus the two dimensional $O(2)$--model 
can have point defects or vortices. 

Kosterlitz and Thouless used approximate renormalization group [RG]
methods to show the existence of a phase transition driven by the 
binding of such vortices in the two dimensional $XY$--model (which 
is also called the $O(2)$ non-linear $\sigma$--model or the two 
component Heisenberg model) at finite non-zero temperature 
\cite{Be71,KT}. The two dimensional $XY$--model remains the generic 
model for the study of Kosterlitz--Thouless [KT] type phase 
transitions. 

The partition function for the classical $n$--component Heisenberg 
model on a lattice $\Lambda$ of linear extent $L$ is
\begin{equation}
 Z_L(\beta,h)
 =
 \int_{S_{n-1}}{
 \prod_{x \in \Lambda}{
                 d{\vec{\sigma}}_x
                 e^{\beta S
                    + {\vec{h}} . {\vec{M}} 
                   }
               }}
\quad ,
\label{pf}
\end{equation}
where
\begin{equation}
 S
 =
 \sum_{x \in \Lambda}
 \sum_{\mu = 1}^{d}
 {\vec{\sigma}}_x .
 {\vec{\sigma}}_{x+\mu}
 \quad ,
 \quad \quad
 {\vec{M}}
 =
 \sum_{x \in \Lambda}
 {\vec{\sigma}}_x
\quad ,
\label{SM}
\end{equation}
and where $\beta = 1/kT$ is the (reduced) inverse temperature ($T$ 
being the temperature and $k$ the Boltzmann constant). The 
$n$--component spins ${\vec{\sigma}}_x$ have unit modulus
and ${\vec{h}}$ is a (reduced) external field. This is the 
infinite bare coupling limit of the $O(n)$ $\phi^4$--model. It is 
believed that the $n$--component Heisenberg model is in the same 
universality class as the general $O(n)$ $\phi^4$--model.

The scenario proposed by Berezinskii \cite{Be71} and Kosterlitz and 
Thouless \cite{KT} is that at temperatures  above some critical 
value ($\beta < \beta_c$) the vortices and antivortices are 
unbounded and serve to disorder the system. The vortex chemical 
potential is a relevant variable. Decreasing the temperature 
($\beta \nearrow \beta_c$) causes the vortices and antivortices 
to bind, thereby decreasing their relevance as dynamical degrees 
of freedom. The model remains critical  (thermodynamic functions 
diverge) for all $\beta > \beta_c$ and the critical exponents are 
dependent on temperature. In terms of the reduced temperature, 
which is defined as
\begin{equation}
  t = 1 - \frac{\beta}{\beta_c}
\quad ,
\label{redt}
\end{equation}
the leading critical behaviour of the correlation length, 
susceptibility  and  the  specific  heat is given in \cite{KT} as
\begin{eqnarray}
 \xi_\infty(\beta) & \sim & e^{bt^{-\nu}} 
 \quad , \label{ktxi} 
 \\
 \chi_\infty(\beta) & \sim & \xi_\infty^{\tilde{\gamma}} 
 \quad , \label{ktchi}
 \\
 C_\infty(\beta) & \sim & \xi_\infty^{\tilde{\alpha}} + 
 {\rm{constant}}
 \quad ,
 \label{ktcv}   
\end{eqnarray}
where for  $t \rightarrow 0^+$,
$\nu = 1/2$, 
$\tilde{\gamma} = \gamma / \nu = 2-\eta_c = 7/4$ and
$\tilde{\alpha} = \alpha / \nu = -d =-2$.


The exponential scaling behaviour of (\ref{ktxi}) is referred to as 
essential scaling to distinguish it from the (more usual) power-law 
behaviour of (\ref{convxi}). Thus the `KT scenario' means a phase 
transition (i) driven by a vortex binding mechanism {\em{and}} 
(ii) exhibiting essential scaling behaviour.

Although essential scaling is also often referred to as KT scaling, 
this (exponential) scenario was known to exist in certain models 
before the seminal work of \cite{KT}. In particular, the BCSOS 
and F- models are special cases of the $6$--vertex model and are 
exactly solvable in the absence of an external field \cite{Baxter}. 
This means that the even thermodynamic functions and in particular 
the free energy and the correlation length are exactly obtainable 
in these models.  The latter has the form (\ref{ktxi}) (also 
with $\nu = 1/2$) and the  singular part of the free energy is 
\begin{equation}
 f^{\rm{BCSOS}}_\infty(\beta) \propto \xi_\infty^{-2}
\quad .
\label{bcsos}
\end{equation}
This model has not been solved in the presence of an external field
and the odd critical exponents ($\gamma$, $\eta$ and $\delta$) are 
not, therefore, rigorously determinable. (Neither are these critical 
exponents determinable from the usual scaling relations since they 
don't all hold in the case of an essential critical point 
\cite{KT,Baxter}).

The reasoning presented by KT supporting an essential scaling 
scenario for the $XY$--model (see also 
\cite{JoKa77,Wi78,AmGo80,KaZi81}) is non--rigorous and a primary 
concern of lattice (non-perturbative) field theory has been its 
verification. An unambiguous verification of all the KT predictions 
has until now however proved elusive. Recently, however, Hasenbusch 
et al. \cite{HaMa94} numerically matched the RG trajectory of  the 
dual of the $XY$--model to that of the BCSOS model. This provides 
further evidence that at least the even critical exponents of the 
$XY$--model are those given by KT.

Most Monte Carlo [MC]  and high temperature expansion [TE] 
analyses of the $XY$--model have concentrated on the numerical 
determination of $\nu$ and $\eta$ from (\ref{ktxi}) and (\ref{ktchi}) 
as they stand. These determinations are notoriously difficult not 
least because multi-parameter fits are required. Commonly, $\beta_c$ 
is determined by fixing $\nu = 1/2$ in (\ref{ktxi}). Typically, 
however, the subsequent fit to (\ref{ktchi}) yields a value 
significantly different from the KT prediction $\eta_c = 1/4$. 
Unconstrained four-parameter fits are hampered by slow convergence 
problems due to extremely shallow valleys in the 4--parameter space 
\cite{GuDe88} and also fail in general to quantitatively and 
unambiguously confirm the KT predictions. Table~\ref{taba} 
demonstrates this by listing some of the more recent 
estimates for $\beta_c$ and $\eta_c$ from Monte Carlo [MC]  and 
high temperature expansion [TE] methods.

\noindent
\begin{table}[ht]
\caption{Estimates for $\beta_c$ and $\eta_c$ 
for the $XY$--model from a selection of 
papers. Those marked with an asterisk 
favour conventional as opposed to 
essential scaling behaviour.}
\label{taba}
\begin{center}
\begin{small}
\vspace{0.5cm}  
\noindent\begin{tabular}{|l|l|l|l|l|} 
    \hline 
    \hline
     Authors & Year  & Method & $\beta_c$ & $\eta_c$  \\
      & & & &  \\
    \hline
Fern\'andez et al. \cite{FeFe86}&1986& MC (FSS)&$1.1$&$0.24(3)$\\
\hline
Seiler et al. 
\cite{SeSt88}\hspace{-2.5mm} $~^*$ &1988&MC&$1.01(1)$&$0.38(2)$
\\ \hline
Gupta et al. \cite{GuDe88} & 1988 & MC & $1.114(2)$  & $0.34$ \\
\hline
Wolff \cite{Wo89} & 1989 & MC  &  $1.12(7)$   & $0.280(4)$ \\
\hline
Biferale, Petronzio \cite{BiPe89} &1989&MC(FSS)&$1.112(2)$&0.243(7)\\
\hline
Butera, Comi,Marchesini \cite{BuCo89}&1989&TE&$1.112(6)$&0.300(25)\\
\hline
Ferer, Mo \cite{FeMo90}&1990& MC(BSRG) & $1.115(10)$   & $0.27(3)$\\
\hline 
Hulsebos, Smit, Vink \cite{HuSm90} & 1990 & MC(FSS) & $1.13$ & \\
\hline 
Edwards, Goodman, Sokal \cite{EdGo91}&1991&MC(FSS)&$1.130(15)$ & \\
\hline 
Gupta, Baille \cite{GuBa91}&1991&MCRG(FSS)&$1.119(5)$&$0.235(5)$ \\
\hline 
Butera, Comi \cite{BuCo93} &1993&TE& $1.118(3)$  & $0.27(2)$ \\
\hline
Janke, Nather \cite{Janke} & 1993 & MC(FSS) &   & $0.2389(6)$ \\
\hline
Catterall, Kogut, Renken \cite{CaKo93}&1993&MC(FSS)& & $0.247(1)$ \\
\hline
Butera, Comi \cite{BuCo94}  & 1994 & TE      &  & $0.27(1)$  \\
\hline
Schultka, Manousakis \cite{ShMa94}&1994&MC(FSS) &$1.119(4)$  & \\
\hline 
Hasenbusch, Marcu, Pinn \cite{HaMa94} &1994&MC(BSRG)&$1.1197(5)$& \\
\hline 
Olsson \cite{Ol95}  & 1995 & MC& $1.1209(1)$ &  \\
\hline
Kim \cite{Kim} \hspace{-3.4mm} $~^*$ & 1995 &MC& $1.108(3)$ & \\
\hline 
Campostrini et al. \cite{CaPe95}&1995&TE&$1.118(4)$&$0.228(1)$
 \\
    \hline
    \hline
  \end{tabular}
\end{small}
\end{center}
\end{table}

In a recent letter \cite{KeIr95} we presented a very general 
theoretical argument showing that the essential scaling scenario 
is not self--consistent unless multiplicative logarithmic 
corrections are included. We believe that the measured values 
of $\eta_c$ in table 1 deviate from the theoretical value 
($\eta_c = 1/4$) because these logarithmic corrections have 
been ignored in these analyses (the only exception being 
\cite{BuCo93}). Our assertion is that it is essential to take 
account of the possibility of multiplicative logarithmic 
corrections in any attempt to numerically verify an essential 
scaling scenario.

In \cite{KeIr95} we also presented a new method involving the 
finite--size scaling [FSS] of the first Lee--Yang zero \cite{LY} 
whereby the critical point $\beta_c$ and the leading and logarithmic 
correction critical indices can be determined numerically from two
parameter (straight line) fits. Here we wish to elaborate on these 
theoretical and numerical approaches and to extend our analysis to 
the first ten Lee--Yang zeroes.

\subsection{The step Model} 

The step model or  sgn  $O(2)$ model is closely  related  to the 
$XY$--model \cite{GuJoTh72,GuJo73,LeSh87,LeSh88}.
Its partition function is given by (\ref{pf}) with 
\begin{equation}
 S^{\rm{step}}
 =
 \sum_{x \in \Lambda}
 \sum_{\mu = 1}^{d} \hbox{sgn}(
 {\vec{\sigma}}_x\cdot
 {\vec{\sigma}}_{x+\mu})
\quad  .
\label{step}
\end{equation}
The question of criticality of this model has, until now, been 
unresolved despite several analyses based on high temperature 
series and on numerical  simulation \cite{NyIr86,SVWi88}. The 
interest in the model arises from its possible membership of the 
Kosterlitz--Thouless universality class. Like the  $XY$--model, 
the step model has a configuration space which is globally and 
continuously symmetric. Unlike the $XY$--model, however, the 
interaction function is discontinuous and the Mermin--Wagner 
theorem \cite{MW} remains unproven for this case (see however 
\cite{PaSe92}). Nonetheless, it is expected that if a phase 
transition exists in the step model, it should not be to a phase 
with long range order \cite{GuJo73,SVWi88,GuNy78}.

The KT phase transition of the $XY$--model is understood to be 
driven by the binding/unbinding of vortices. The energetics of 
vortex formation in the step model are, however, very different 
from the $XY$--model \cite{LeSh88,SVWi88,GuNy78}. Since vortices 
with effectively zero excitation energy can be created at all 
non-zero temperatures, the usual KT argument does not naturally 
lead one to expect such a phase transition in the step model.

S\'anchez--Velasco and Wills \cite{SVWi88} presented evidence of 
critical behaviour starting at $\beta_c^{\rm{step}}=0.91\pm0.04$. 
This was based on FSS of the spin susceptibility. Since the 
associated critical index $\eta(\beta_c^{\rm{step}})$ was 
significantly greater than that measured for  the $XY$--model, 
it was concluded that the step and $XY$ models are not in 
the same universality class. In this paper we present evidence that 
the step model is {\em not} in fact critical at that temperature.
However it {\em is} critical at lower temperatures with a critical 
index $\eta(\beta_c^{\rm{step}})$ compatible with the $XY$ value. 

These results are consistent with the possibility that the $XY$-- 
and step models belong to the one universality class. Since the KT 
vortex unbinding mechanism is not believed to be responsible for 
the phase transition of the step model, the same scenario has to 
be questioned in the $XY$--model. This is not the first time the 
KT scenario has been questoned. For earlier counter-evidence to 
both the physical KT picture as well as the quantitative essential 
scaling predictions  see \cite{Zi76,LuSc77,FuSo84} and 
\cite{SeSt88,PaSe88}. Further criticisms of the conventional 
view are found in \cite{PaSe95}.

The accuracy afforded by the Lee--Yang zeroes study and the 
consideration of multiplicative logarithmic corrections are 
essential for  the present analysis \cite{IrKe95}.

\section{Lee--Yang Zeroes and Logarithmic Corrections to Scaling}

\setcounter{equation}{0}

The subject of partition function zeroes was introduced in 1952 by 
Lee and Yang \cite{LY} as an alternative way to understand the onset 
of criticality in statistical physics models. For a finite system 
the zeroes of the partition function are strictly complex 
(non--real). As $L$ is allowed to go to infinity one generally 
expects these zeroes to condense onto smooth curves. Zeroes in the 
plane of complex external magnetic field $h$ are refered to as 
Lee--Yang zeroes. Lee and Yang further showed that for certain 
Ising--type systems these zeroes are in fact restricted to the 
imaginary $h$ axis (the Lee--Yang theorem) \cite{LY}. Dunlop and 
Newman proved that the Lee--Yang theorem holds for the two 
dimensional $XY$--model \cite{DuNe75}. While no specific proof 
exists for the step model, its similarities to the $XY$--model 
lead one to anticipate a similar locus of zeroes. In practice, 
one obtains numerical confirmation of this hypothesis by  
explicit determination of the locus.

In the symmetric phase, $t>0$, the Lee--Yang zeroes lie away from 
the real $h$-axis, pinching it only as $t \searrow 0$ (in the 
thermodynamic limit). For systems obeying the Lee--Yang theorem 
this pinching occurs at $h=0$, prohibiting analytic continuation 
from ${\rm{Re}}(h) < 0$ to ${\rm{Re}}(h) > 0 $.
                    

\subsection{Lee--Yang Zeroes}

Consider a partition function of the form
\begin{equation}
 Z_L(\beta,h)
 =
 \frac{1}{{\cal{N}}_L}
 \sum_{\{\vec{\sigma}_x\}}{
                 e^{\beta S
                    + h {\hat{n}}.{\vec{M}}
               }}
\quad ,
\label{pf1}
\end{equation}
where ${\hat{n}}$ is a unit vector defining the direction of the 
external magnetic field and $h$ is a scalar parameter representing 
its strength. The factor ${{\cal{N}}_L}$  ensures that 
$Z_L(0,0) = 1$. Here $S$ and ${\vec{M}}$ are given by (\ref{SM}) 
or (\ref{step}), $S$ taking real values from the interval 
$[-dL^d,dL^d]$ ($dL^d$ is the number of links in  $\Lambda$) 
and $M = (\hat{n} . \vec{M})$ taking the real values in  
$[-L^d,L^d]$ ($L^d$ being the number of sites in $\Lambda$).
 
Suppose now that the $M$--range is binned such that there are 
$2N_M$ $M$--bins of width $\Delta_M  =  {L^d}/{N_M}$.
Defining the integrated spectral density 
$\rho_{L,\beta}^{(\Delta)}(M)$ 
as the number of
configurations having given $M$-- values,
the partition function for this system can be written as
\begin{equation}
 Z_L^{(\Delta)}
 (\beta ,h)
 =
 \frac{e^{ -h M_{\rm{max}}  }}{{\cal{N}}^{(\Delta)}_L}
 \sum_{k=0}^{2 M_{\rm{max}} / \Delta_M}{
          \rho_{L,\beta}^{(\Delta)}\left(
                                    k \Delta_M - M_{\rm{max}}
                                 \right)
                                 \left(
                                    e^{h\Delta_M}
                                 \right)^k
}
\quad ,
\label{fouriert}
\end{equation}
where $M_{\rm{max}}=(N_M-\frac{1}{2})\Delta_M$ and where 
$k \in {\mmath N}$ and ${\cal{N}}^{(\Delta)}_L$ is such that 
$Z_L^{(\Delta)}(0,0)=1$. Thus $Z_L^{(\Delta)}(\beta,h)$ is 
proportional to a polynomial  with real coefficients of order
$2 M_{\rm{max}} / \Delta_M$ in the `fugacity' $z = \exp{(h \Delta_M)}$
and as such has $2 M_{\rm{max}} / \Delta_M$ zeroes in the complex
$z$--plane. 

For the Ising model (for example) $\Delta_M$ is the minimum
change in the total magnetization upon flipping a single spin. 
There $\Delta_M = 2$. In the $XY$--model the $M$--range is 
continuous, even for a finite system. However, when we come 
to the numerical analysis of the $XY$--model, the Monte Carlo 
integration method can only sample a discrete subset of the 
infinite range of configurations open to the system. It thus 
accesses only a discrete sample of $M$--values from the range 
$[-L^d,L^d]$. If no binning is used in the numerical approach, 
then the minimum separation between two such $M$--values will 
play the r\^ole of $\Delta_M$ above. In any case, the partition 
function is still proportional to a polynomial in the fugacity.
The real and imaginary parts of the  partition function 
are separately periodic in ${\rm{Im}}(h)$ with period
$2\pi / \Delta_M $. The pattern of Lee--Yang zeroes is 
itself therefore  periodic in ${\rm{Im}}(h)$
with the same period.

\subsection{Scaling and Corrections to scaling}

Assume that the Lee--Yang theorem holds for the binned system. 
In terms of its zeroes,
\begin{equation}
 z_j(\beta,L) = e^{i \Delta_M \theta_j(\beta,L)}
\quad ,
\label{uzero}
\end{equation}
the partition function (\ref{fouriert}) is 
\begin{equation}
 Z_L^{(\Delta)}(\beta,h)
 =
 \frac{e^{-h M_{\rm{max}}}}{{\cal{N}}^{(\Delta)}_L}
 \rho_{L,\beta}^{(\Delta)}\left( M_{\rm{max}} \right)
 \prod_{j=1}^{2 M_{\rm{max}} / \Delta_M}
 \left( z - z_j(\beta,L)  \right) 
 \quad .
\label{zitou}
\end{equation}
Assume
\begin{equation}
 f_L(\beta,h)
 =
 \lim_{\Delta_M \rightarrow 0}{
 f_L^{(\Delta)}(\beta,h)
 } 
 =
 \lim_{\Delta_M \rightarrow 0}{
 L^{-d}                          
 \ln{Z_L^{(\Delta)}(\beta,h)}
 }
 \quad .
\label{hope}
\end{equation}
Zeroes on the unit circle in the complex fugacity plane correspond 
to zeroes on the imaginary $h$ axis. Following 
\cite{Abe,SuzukiLY,KeLa94}, we define the density of these zeroes as
\begin{equation}
 g_L(\beta,\theta) 
 =
 \frac{dG_L(\beta,\theta)}{d\theta}
 =
 L^{-d}
 \sum_{j=1}^{2 M_{\rm{max}}}{\delta(\theta - \theta_j(\beta,L))}
\label{gG}
\quad .
\end{equation}

\paragraph{Scaling in the Thermodynamic Limit}
Using the symmetry and periodicity \cite{KeIr95} of the pattern of 
zeroes one finds for the the free energy in the thermodynamic 
limit \cite{Salmhofer}
\begin{equation}
 f_\infty (\beta,h)
 =
 f_{\rm{reg}} (\beta,h)
 -
 2 
 \int_{\theta_c}^\infty 
 {\frac{\theta}{h^2 + \theta^2} G_\infty(\beta,\theta)d\theta}
 \quad .
\label{free}
\end{equation}
where $\theta_c(\beta)$ is the Yang--Lee edge
and is given by \cite{GaMi67,KoGr71}
\begin{equation}
 g_\infty (\beta,\theta)=0 
 {\rm{~ ~ for ~ ~}} - \theta_c(\beta) < \theta < \theta_c(\beta)
\quad .
\end{equation}
The zero field susceptibility is the second derivative of the free 
energy  with respect to $h$,
\begin{equation}
 \chi_\infty(\beta)
 = \left. \frac{\partial^2 f_\infty}{\partial h^2}\right|_{h=0}=
 4  \int_{\theta_c(\beta)}^{\infty}
 \theta^{-3}
 G_\infty(\beta,\theta)
 d \theta
\quad .
\label{zerofieldX}
\end{equation}
Following \cite{Abe,SuzukiLY,KeLa94, KeIr95}, the cumulative density 
of zeroes can be written in terms of the susceptibility and the 
Yang--Lee edge:
\begin{equation}
 G_\infty(\beta,\theta)
 =
 \chi_\infty (\beta)
 {( \theta_c(\beta)) }^2
 \Phi \left( \frac{\theta}{\theta_c(\beta)} \right)
\quad ,
\label{twoPhi}
\end{equation}
$\Phi (x)$ being some function of $x$ with
$\Phi(\mid x\mid\leq 1)=0$. 
 
The (singular part of the) specific heat is then
\begin{equation}
C^{\rm{sing}}_\infty(\beta) =  \left.
 \frac{\partial^2f^{\rm{sing}}_\infty(\beta,h)}{\partial \beta^2}
 \right|_{h=0}
=
 -2\int_{\theta_c(\beta)}^\infty
 \theta^{-1}
 \frac{d^2G_\infty\left(\beta,\theta \right)}{d\beta^2}
 d\theta
 \quad .
\label{directerthanAbe}
\end{equation}

The above formulae are quite general and hold for any model provided
only that it obeys the Lee--Yang theorem. To proceed from here we 
have to insert the (model--specific) critical behaviour. Instead 
of the conventional KT formula (\ref{ktchi}) and (\ref{ktcv})
assume, now, the following more general
behaviour for the zero field 
susceptibility, the edge and for the singular part of the specific 
heat
\begin{eqnarray}
 \chi_\infty (\beta) & \sim & \xi_\infty^{\tilde{\gamma}} 
                           (\ln{\xi_\infty})^{\tilde{r}}
\quad ,
\label{KTchi7}
\\
\theta_c (\beta) & \sim & 
\xi_\infty^{\lambda}(\ln{\xi_\infty})^{\tilde{p}}
\quad ,
\label{jjj}
\\
C^{\rm{sing}}_\infty (\beta)
                  &\sim &
 \xi_\infty^{\tilde{\alpha}}
                       (\ln{\xi_\infty})^{\tilde{q}}
\quad ,
\end{eqnarray}
and
\begin{eqnarray}
 \gamma & = & \nu \tilde{\gamma}
 \quad \quad \quad \quad
 r = - \nu \tilde{r}
\quad
\\
 \alpha & = & \nu \tilde{\alpha}
 \quad \quad \quad \quad
 q = - \nu \tilde{q}       
\quad .
\end{eqnarray}
Then (\ref{twoPhi}) and (\ref{directerthanAbe}) give
\begin{equation}
 \xi_\infty^{\tilde{\alpha}-\tilde{\gamma}-2\lambda}
 (\ln{\xi_\infty})^{\tilde{q} - \tilde{r} - 2\tilde{p}}
 \propto 
 \left[
       c_1\frac{\ddot{\xi}_\infty}{\xi_\infty}
       +
       c_2 \left( \frac{\dot{\xi}_\infty}{\xi_\infty} \right)^2
 \right]
 \left\{
         1 + O \left(\frac{1}{\ln{\xi_\infty}}\right)
 \right\}
\quad .
\label{master}
\end{equation}
Inserting the leading KT behaviour for the correlation 
length\footnote{Including logarithmic corrections in (\ref{ktxi}) 
so that $\xi_\infty \sim \exp{(bt^{-\nu})} t^w$ only delivers extra 
additive corrections and has no consequence for the other critical 
indices. Although such corrections may be present in the 
correlation length, $w$ remains undetermined.
} 
(\ref{ktxi}) gives
\begin{equation}
 \tilde{\alpha} = \tilde{\gamma} + 2 \lambda
 \quad ,
\end{equation}  
and
\begin{equation}
 \tilde{q} = \tilde{r} + 2 \tilde{p} + 2\left(1 + 
\frac{1}{\nu}\right)
\quad .
\label{prob}
\end{equation}
Here we have the first indications that  KT scaling behavoiur
(\ref{ktxi}) -- (\ref{ktcv}) is not self--consistent without
multiplicative logarithmic corrections. If there are no 
logarithmic corrections then $\tilde{q} = \tilde{r} = \tilde{p} = 0$ 
and (\ref{prob}) cannot hold.

\paragraph{Finite--Size Scaling}
The finite--size scaling (FSS) hypothesis, first formulated in 1971
by Fisher and co-workers \cite{Fi72}, is a relationship
between the critical behaviour of thermodynamic quantities in the 
infinite volume limit and the size dependency of their finite volume 
counterparts. The general statement of FSS, which is expected to hold 
in all dimensions \cite{KeLa93} is that if $P_L(\beta)$ is the value 
of some thermodynamic quantity $P$ at inverse temperature (coupling) 
$\beta$ measured on a system of linear extent $L$, then
\begin{equation}
 \frac{P_L(\beta_c)}{P_\infty (\beta)}
 =
 {\cal{F}}_P
 \left( 
        \frac{\xi_L(\beta_c)}{\xi_\infty (\beta)}
 \right)
\quad ,
\label{FSS}
\end{equation}
where $\xi_L (\beta)$ is the correlation length of the finite--size 
system. Here ${\cal{F}}_P$ is some $P$--dependent function of the 
scaling  variable $x = \xi_L(\beta_c) / \xi_\infty (\beta)$.
Fixing the scaling variable, one has
\begin{equation}
 \xi_\infty (\beta) 
 \sim
 x^{-1} \xi_L (\beta_c)
\quad .
\end{equation}
Luck has shown that for the two dimensional $XY$ model, 
$\xi_L(\beta_c)$ is proportional to $L$ \cite{Lu82}.
Therefore FSS for the susceptibility and the Yang--Lee edge is
from (\ref{KTchi7}) and (\ref{jjj})
\begin{eqnarray}
 \chi_L (\beta_c) & \sim & L^{\tilde{\gamma}}
                           (\ln{L})^{\tilde{r}}
\label{chifss}
 \\
 \theta_1 (\beta_c) & \sim & L^{\lambda}
                           (\ln{L})^{\tilde{p}}     
\label{edgefss}
\quad .
\end{eqnarray}
Now, from (\ref{zitou}) and (\ref{free}) the magnetic susceptibility 
for a  finite--size system is
\begin{equation}
 \chi_L (\beta) 
 =
 \left.
    \frac{\partial^2 f_L}{\partial h^2}
 \right|_{h=0}
 =
 -
 L^d
 \sum_j{
          \frac{1}{\theta_j^2}
}
\quad .
\label{as1}
\end{equation}
If the lowest lying zeroes exhibit the same FSS behaviour \cite{IPZ}, 
then
\begin{equation}
 \chi_L (\beta) \sim - L^{-d} \theta_1(\beta)^{-2}
\quad .
\label{as2}
\end{equation}
We provide numerical justification for this assumption in 
Sec.~{\ref{4.2}}. Together, (\ref{chifss}), (\ref{edgefss}) 
and (\ref{as2}) give
\begin{eqnarray}
 \lambda & = & - \frac{1}{2}(d+\tilde{\gamma})
\quad , 
\label{l1}
\\
 \tilde{p} & = & - \frac{1}{2}\tilde{r}
\quad ,
\label{p1}
\end{eqnarray}
whence
\begin{equation}
 \tilde{\alpha} = -d
 \quad ,
\label{nojosephson}
\end{equation}
and
\begin{equation}
 \tilde{q} = 2\left(1 + \frac{1}{\nu}\right)
\quad .
\end{equation}
Thus the KT result for the leading specific heat index $\alpha$ is 
recovered. However the correction index is $q = 2(\nu + 1) = 3 $, 
a non-trivial result. There are therefore multiplicative logarithmic 
corrections to the  leading (KT) scaling behaviour of the singular 
part of the specific heat. 
\begin{equation}
 C^{\rm{sing}}_\infty 
  \sim 
 \xi_\infty^{-2} 
 (\ln{\xi_\infty})^6
  \sim
 \xi_\infty^{-2}
 t^{-3}
 \quad . 
\label{cwlog}
\end{equation}
The full specific heat has another (constant) term coming from the 
regular  part of
the free energy and since the leading critical index is negative 
any numerical verification of (\ref{cwlog})
is rendered very difficult if not impossible \cite{Ba83}.

Integrating (\ref{cwlog}) twice with
respect to  $t$ gives the singular part of the free 
energy to be $f_\infty \sim \xi_\infty^{-2}$ as in (\ref{bcsos}). 
Thus we have indirectly verified hyperscaling for the 
two-dimensional $XY$--model and we have done this using 
the self-consistency of the essential scaling behaviour 
and finite--size scaling\footnote{
Hyperscaling and FSS are distinct hypotheses, the latter being the
stronger. In the upper critical dimension ($d=4$) FSS has been
shown to hold where hyperscaling fails \cite{KeLa93}.
}.

The above analytic considerations have yielded no information on
the odd correction exponent $r$. 
The original renormalisation group analysis of Kosterlitz and 
Thouless \cite{KT}, in fact,
implicitly contained the prediction 
\beq
r=-1/16
\label{eq:r16}
\eeq
as noted by \cite{BuCo93}. Subsequent analyses have concentrated on 
the $r=0$ form of the scaling behaviour (\ref{chifss})
and the verification that $\eta(\beta_c)=1/4$.
Allton and Hamer \cite{AlHa88} have conjectured that the
deviation of their determination of $\eta$ from $1/4$ might
be due to logarithmic corrections.

With $\nu = 1/2 $ and $\lambda = -2 + \eta_c / 2$ the FSS formula 
for the first  Lee--Yang zero is 
\begin{equation}
 \theta_1 (\beta_c) \sim L^{\lambda} (\ln{L})^r
\quad .
\label{testr}
\end{equation}
The study of this full scaling form and the numerical 
determination of $r$ is the subject of what follows.

\section{Numerical Results}

 The Monte Carlo (MC) method is a stochastic method of importance 
sampling  and an intrinsically non--perturbative approach
to the calculation of path integrals and expectation values. In this 
section  we wish to report on an application to the $XY$--model in an
attempt to test independently the above scaling scenario.

In recent years the development of more efficient techniques has 
enhanced  the quality and practicality of certain numerical 
calculations  in bosonic  spin systems. These improvements include

\noindent
$\bullet$
the advent of non--local algorithms \cite{Wo89,SwWa87} which greatly 
reduce critical slowing down,

\noindent
$\bullet$
the rediscovery and development of the spectral density method which 
allows extrapolation away from the simulation point \cite{Ma84,FeSw88},

\noindent
$\bullet$
the use of multihistograms to extend the coupling  range over which 
extrapolation of information is reliable \cite{Bo89,KaKa91}.

The aim of the present section is to present numerical evidence
for the existence (or otherwise) of logarithmic corrections to 
scaling. We shall adopt a self-consistent strategy.
\begin{enumerate}
\item
 \begin{enumerate}
 \item Temporarily ignore the effect of logarithmic corrections 
and aim to 
 extract the basic critical parameters $\beta_c$ and $\eta(\beta_c)$. 
 One verifies, from the scaling of $\chi_L(\beta)$, the existence
 of a critical region below some approximately determined 
 temperature $T_c \equiv 1/ \beta_c$ 
 and that the effective values of $\tilde\gamma$ near this
 temperature include the expected KT value  $2- \eta_c = 7/4$.
 \item {\em Assuming} this value holds at $\beta_c$, use it to 
extract
 the so-called Roomany-Wyld beta-function approximant (\ref{eq:RWA3})
which is then valid only {\em at} the critical point.
 \item Deduce $\beta_c$ from the zeroes of this function using the
residual $L$ (finite--size) dependence to estimate systematic errors.
 \end{enumerate}
\item
 \begin{enumerate}
 \item Now use multi-histogramming techniques 
to study the scaling of Lee--Yang zeroes 
(with much higher precision) over
a range of candidate $\beta_c$ values around the above best 
estimate.
 \item Verify that a critical region exists and that the leading
behaviour is compatible with the KT value ($\lambda=-15/8$)
 \item Use the quality of the scaling fits to 
 $\theta_1(\beta_c)$ of (\ref{testr}) at each candidate $\beta_c$
 to establish the possible existence of logarithmic corrections.
 \end{enumerate}

\end{enumerate}

\subsection{Simulation Details}

A non--local algorithm based on that of Wolff \cite{Wo89} was used 
to  simulate the $XY$-- and step models at zero magnetic field 
($h=0$) on square lattices of sizes $L=32,64,128$  and  $256$. 
In the case of the step model, additional simulations were 
performed at $L=48$. The values of $\beta$ at which the  simulations 
took place ($\beta_o$) were evenly spaced in steps of $0.02$ between 
$0.90$ and $1.20$ for the $XY$--model and with various degrees of 
spacing between $0.86$ and $1.40$ for the step model. 
At each simulation point, $100,000$ measurements 
of $S$ and $\vec{M}$ were made using (\ref{SM}) and  (\ref{step}).

In these isotropic two component models (with no external field) 
the orientation of the magnetization is not fixed in any direction.
One  uses for the magnetization its magnitude
$M = \sqrt{M_x^2 + M_y^2}$ where $M_x$ and $M_y$ are the $1$- and 
$2$-components 
of the magnetization in the two dimensional internal spin space 
\cite{BiSt87}.

\subsection{Approximate Critical $\beta$ Determination}

The expected finite--size scaling behaviour of critical systems can 
be used to locate critical points and determine critical parameters
of an infinite system.
A convenient method for accomplishing this \cite{Ni76, RoWy80} is
to extract an approximation to the Callan-Symanzik beta
function \cite{CaSy70}
from the size dependance (in lattice units)
of some physical quantity.

Consider the measurement of a physical quantity ${\cal Q}$ 
on a lattice of linear extent ${\cal L}=La$ where $a$ is the 
lattice spacing in physical units. The measurement of ${\cal Q}$ 
on the lattice at bare (dimensionless) coupling $\beta$ will result 
in some number expressed in lattice units: $Q_L(\beta)$.
Renormalisation may be viewed \cite{Ni76} as 
varying the scale unit $a$ while maintaining physical 
quantities such as ${\cal L}$ and ${\cal Q}$ fixed
at their physical (bulk) values. Thus, the lattice 
granularity $L={\cal L}/a$ varies when $a$ does. Fixed physics 
then requires that the bare coupling $\beta$ is correspondingly
tuned and the measure of its response is just the Callan Symanzik 
[CS] beta function 
\beq
 B(\beta)=-\mu \beta^{-2}
{{\partial \beta}\over{\partial\mu}},\qquad\mu\equiv{1\over a}\, .
\label{eq:B0}
\eeq

\paragraph{Roomany-Wyld approximant}
Roomany and  Wyld  \cite{RoWy80} showed how, 
for ${\cal Q}\equiv {\cal{M}}=1/\xi$ (the mass-gap), one
can readily extract an estimate of $B(\beta)$ from finite 
lattice measurements of 
$Q_L(\beta)\equiv m_L(\beta) = 1/ \xi_L(\beta)$.
The generalisation to other physical quantities was made by
Hamer and Irving \cite{HaIr}.

To find a preliminary estimate for $\beta_c$
(step 1(a) above), we ignore logarithmic corrections 
and write FSS for  the magnetic susceptibilty
(from (\ref{chifss})) as
\beq
\chi_L^{\rm{approx}}\sim L^{{\gamma}\over{\nu}}\, .
\label{eq:chiS1}
\eeq
We consider the physical, dimensionless combination 
\beq
{\cal P}\equiv {\chi}{\cal L}^{{\gamma}\over{\nu}}=
L^{{\gamma}\over{\nu}}{\chi}a^{{\gamma}\over{\nu}}=
L^{{\gamma}\over{\nu}}\chi_L(\beta)\, .
\label{eq:Phys}
\eeq
The corresponding CS equation is
\beq
d{\cal P}={{\partial{\cal P}}\over{\partial \beta}}d\beta +
{{\partial {\cal P}}\over{\partial L}}dL
=0
\label{eq:CS}
\eeq
where $L$ is treated as a continuous variable, for the present.
The corresponding beta function is then deduced as
\beq
B(\beta)=\beta^{-2}{{\partial\ln P}\over{\partial\ln L}}/
{{\partial\ln P}\over{\partial \beta}}\, .
\label{eq:B1}
\eeq
Numerical approximations to the derivatives are then applied
as circumstances dictate.  Using lattice sizes $L$ and $L'$
to estimate the logarithmic derivative one obtains the generalised
RW approximant
\beq
B_Q(t,L,L')=
{{
\phi_Q - \ln[Q_{L^\prime}(t)/Q_L(t)]/\ln(L'/L)
}\over{
{1\over 2}
{{\partial}\over{\partial t}}\ln [Q_{L^{\prime}}(t)Q_L(t)]
}}
\label{eq:RWA3}
\eeq
Classical numerical interpolation may be used to estimate
the derivative appearing in the denominator. 

\paragraph{Application to the XY--Model}
The present application is to the determination of the critical 
point using finite lattice data for the susceptibility
alone (ignoring multiplicative logarithmic corrections). 
Following through the above arguments for the expected 
beta function critical behaviour, one finds
\beq
B(t)\approx {{t^{\nu+1}}\over{b\nu}}\, .
\eeq
This is independent of $\gamma$ but the construction of
the numerical approximant (\ref{eq:RWA3}) requires its knowledge
($\phi_Q=\gamma/\nu$). There are two further complications
\begin{itemize}
\item the presence of logarithmic corrections has been ignored up 
to this point
\item there 
are several arguments to show that $\tilde\gamma\equiv 2-\eta$ is
weakly dependent on $t$. The KT prediction is that, 
at $t=0$ ($\beta = \beta_c$), $\eta =\quarter$.
\end{itemize}
Thus the RW beta function will only be even approximately valid at 
the
location of the critical point. Since it should vanish there, this
is good enough to locate the critical point subject to the assumed
value of the index $\tilde\gamma$.

\paragraph{Results}
The above has been carried out for the XY--model 
using lattice sizes $L=32,64,128$ and
$256$. Specifically, the effective $\tilde{\gamma}(T,L)$ has been
extracted from 3 pairs of consecutive
$L$ values using the scaling implied by (\ref{eq:chiS1}).
The difference between effective
values was monitored and seen to vanish above $\beta\equiv 1/T\simeq
1.1$. The average values were also montitored and seen to pass through
$1.75$ at the same place. 
Statistical errors were estimated assuming that
those of the input $\chi_L(\beta)$ data are Gaussian and independent.

\begin{figure}[htb]
\vspace{13cm}               
\includegraphics{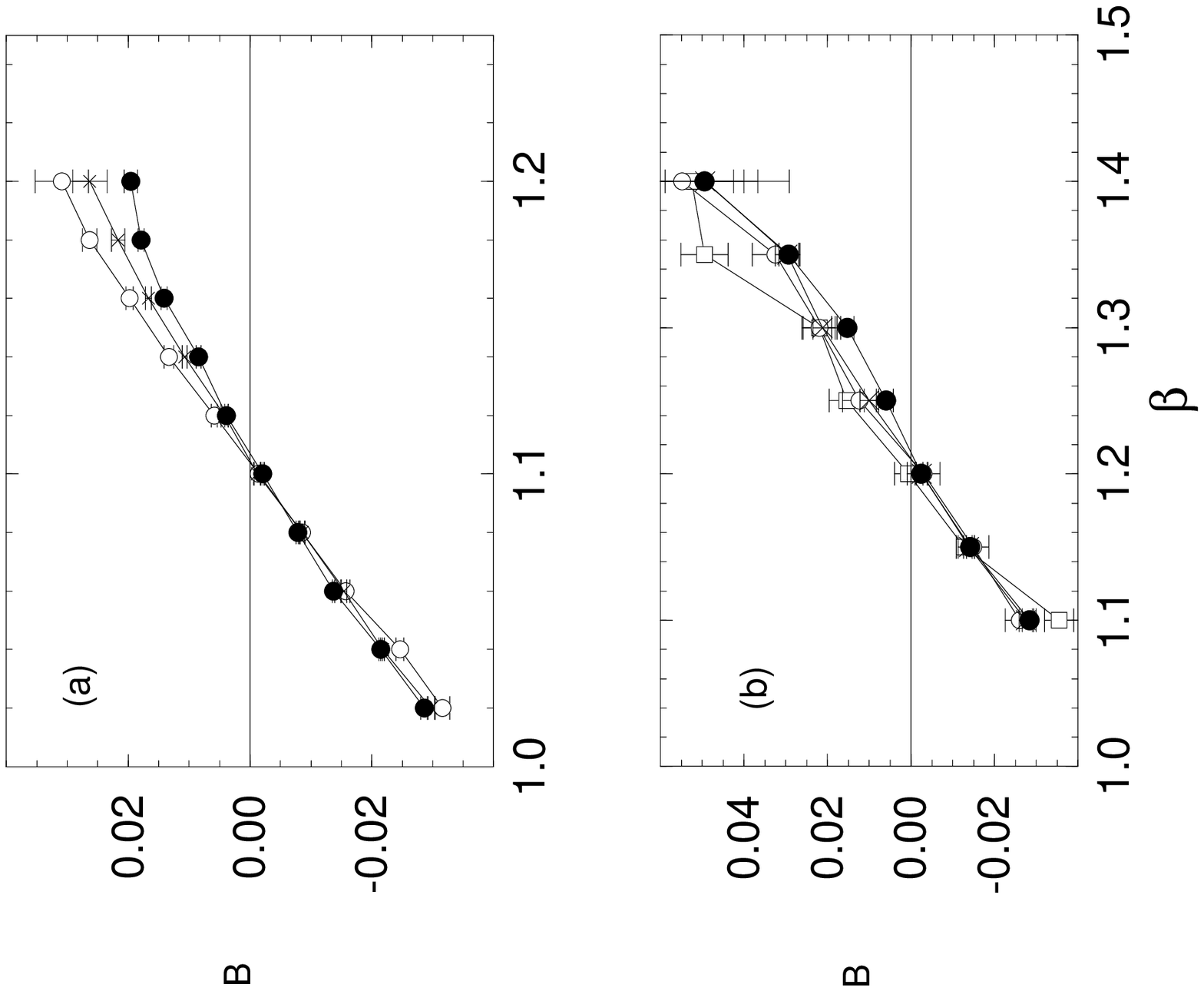}
\caption{The Roomany--Wyld beta function approximants for
(a) the $XY$--model and (b) the step model. The symbols 
\footnotesize $\Box$\normalsize, 
$\circ$, $\times$ and $\bullet$} are in order of increasing
lattice size.
\label{fig01:rw}
\end{figure}

Next, treating $\tilde{\gamma}=1.75$ as constant in the neighborhood
of $\beta=1.1$, the RW approximant to the beta function was calculated
for consecutive pairs of available $L$ values. As expected, 
these vanished at  stable values of $\beta$ as  shown 
in Fig.~\ref{fig01:rw}. The estimates of critical $\beta$, 
which we now denote $\beta_{\rm{RW}}$,
were determined by numerical interpolation
and statistical errors determined as above. The values are
shown in Table~\ref{tab:RWbc}.

\noindent 
\begin{table}[ht]
\caption{
Preliminary $\beta_{\rm{RW}}$ estimate from RW approximant zeroes.
}
\label{tab:RWbc}
\vspace{0.5cm} 
\begin{center}
\noindent\begin{tabular}{|c|c|c|c|c|}
    \hline
    \hline
  & \multicolumn{2}{c|}{}
             & \multicolumn{2}{c|}{} \\
  & \multicolumn{2}{c|}{$XY$--Model}   
             & \multicolumn{2}{c|}{step Model} \\
    \hline
 $L/L'$ & $ \beta_{\rm{RW}} $ & stat. err.& 
$ \beta_{\rm{RW}}^{\rm{step}} $ & stat. err.
 \cr
 & & & &
\cr
    \hline
32/48  &   ---  &  --- & 1.1977 & 0.0095
\cr
    \hline
48/64  &   ---  &  --- & 1.2089 & 0.0094
\cr
    \hline
32/64  &   1.1040  &  0.0014 & --- & ---
\cr
    \hline
64/128  &   1.1045  &  0.0014 & 1.2104 & 0.0052
\cr
    \hline
128/256  &   1.1064  &  0.0011 & 1.2184 & 0.0068
\cr
    \hline
    \hline
\end{tabular}
\end{center}
\end{table}

The spread of values with the choice of $L,L^\prime$ used to 
determine them is clearly very small and can be used to estimate 
a systematic error (within the approximation of ignoring 
multiplicative logarithmic corrections).
We estimate
\beq
\beta_{\rm{RW}}=1.106\pm .002\pm .002\, .
\eeq
where the errors are respectively statistical and systematic.

We emphasise that we ignore logarithmic corrections in this section
and $\beta_{\rm{RW}}$ is therefore only a preliminary or approximate 
estimate for $\beta_c$.

\paragraph{Application to the Step Model}
A similar analysis was performed for the step model 
using lattice sizes $L=32,48,64,128$ and $256$. 
The RW approximant results are also shown in Table~2
and the corresponding $\beta_{\rm{RW}}^{\rm{step}}$ 
estimates in Fig.~1b. 
Although the data are somewhat noisier, the behaviour
of the step model is strikingly similar to that of the XY-model. 
The preliminary estimate of critical $\beta$ is
\beq
\beta^{\rm{step}}_{\rm{RW}}=1.21\pm 0.01\pm 0.01\, .
\eeq

\subsection{Lee--Yang Zeroes and Numerical Test of Scaling Scenario}

\subsubsection{Determination of Lee--Yang Zeroes}

When the external field
is complex ($h = h_r + i h_i $), 
the partition function (\ref{pf}) can be rewritten
\begin{equation}
  Z_L(\beta,h_r + i h_i)
  =
  {\rm{Re}} Z_L(\beta,h_r + i h_i)
  + i
  {\rm{Im}} Z_L(\beta,h_r + i h_i)
\, ,
\end{equation}
where
\begin{eqnarray}
 {\rm{Re}} Z_L(\beta,h_r + i h_i)
 & = &
 \frac{1}{\cal{N}}
 \sum_{\{{\vec{\sigma}}_x\}}{e^{\beta S + h_r M} \cos{(h_i M)}}
\, ,
\nonumber
\\
 ~
 & = &
 Z_L(\beta,h_r)
 \langle
 \cos{(h_i M)}
 \rangle_{\beta,h_r}
\label{ReZ}
\, ,
\end{eqnarray}
and
\begin{eqnarray}
 {\rm{Im}} Z_L(\beta,h_r + i h_i)
 & = &
 \frac{1}{\cal{N}}  
 \sum_{\{{\vec{\sigma}}_x\}}{e^{\beta S + h_r M} \sin{(h_i M)}}
\, ,  
\nonumber
\\
 ~
 & = &
 Z_L(\beta,h_r)
 \langle
 \sin{(h_i M)}
 \rangle_{\beta,h_r}
\label{ImZ}
\, .
\end{eqnarray}
According to the Lee--Yang theorem the zeroes are on the imaginary
$h$--axis ($h_r = 0$) where
$ \langle \sin{(h_i M)} \rangle_{\beta,h_r}$ vanishes and so the 
Lee--Yang zeroes are simply the zeroes of
\begin{equation}
 {\rm{Re}} Z_L(\beta,i h_i)
 \propto
  \langle
 \cos{(h_i M)}
 \rangle_{\beta,0}
\, .
\end{equation}
Thus the Lee--Yang zeroes are easily  found
and at no stage is a simulation with complex action involved.

Errors were calculated using the straightforward
bootstrap method where the data for each $\beta_o$ 
are resampled $N_{\rm{boot}}= 100$, times (with replacement) 
leading to $N_{\rm{boot}}$ estimates for $\theta_1$,
from which the variance and bias can be calculated.
In fact, to take account of any correlation present in the 
100,000 values of $S$ and $M$ at each $\beta_0$, the following 
method of error determination was (also) used.

At each $\beta_0$ the 100,000 data are split into $N_{\rm{split}}$ 
subsets each containing $100,000/N_{\rm{split}}$ data. The bootstrap 
method  is then applied to each of these $N_{\rm{split}}$ subsets 
using  $N_{\rm{boot}} = 50$ in each case. 
If the data is strongly correlated or if the error is strongly 
dependent on whatever correlation is present, the dependence 
of the error on $N_{\rm{split}}$ should be visible.
In fact, plots of the relative error in the position of the zero 
against $N_{\rm{split}}$ at each $\beta_0$ and at each $L$ reveal 
little or no $N_{\rm{split}}$ dependency. Thus the data are highly 
uncorrelated and the
error estimates reliable.
Moreover, within the range of $\beta_0$--values studied, there 
appears to be only a weak dependence of the relative  errors on 
$\beta_0$ (the relative errors decrease slightly with increasing 
$\beta$ as the zero approaches the real axis) and none discernable 
on $L$. For the $XY$--model these relative errors are 0.0003 
for the lowest lying zeroes, increasing to 0.0020 for the 15th 
zeroes while for the step model  they increase from 0.0004
to 0.0030.

\paragraph{Multihistograms and the Spectral Density Method}
In order to be able to find the Lee--Yang zeroes at temperatures
other than the simulation points $\beta_0$, we employ a 
multihistogram reweighting technique (the spectral density method 
\cite{Ma84,FeSw88,Bo89,KaKa91}) 
which accomodates extrapolation between $\beta_0$ values.
In view of the large amounts of data now involved it is  
necessary to introduce binning. For the  spectral density 
$\rho_L^{(\Delta)}(S,M)$ Sec.2.1 we bin the raw
histograms in a $500 \times 100$ array. 
The results for the Lee--Yang zeroes turn out to be very stable
with respect to the bin sizes. Altering the number and size of bins
has only a tiny effect on the position of
the zero (well within the eventual errors). 

\subsubsection{The $XY$--Model}
 
\paragraph{Scaling of Lowest Lee--Yang Zeroes}

\begin{figure}[htb]
\vspace{13cm}
\includegraphics{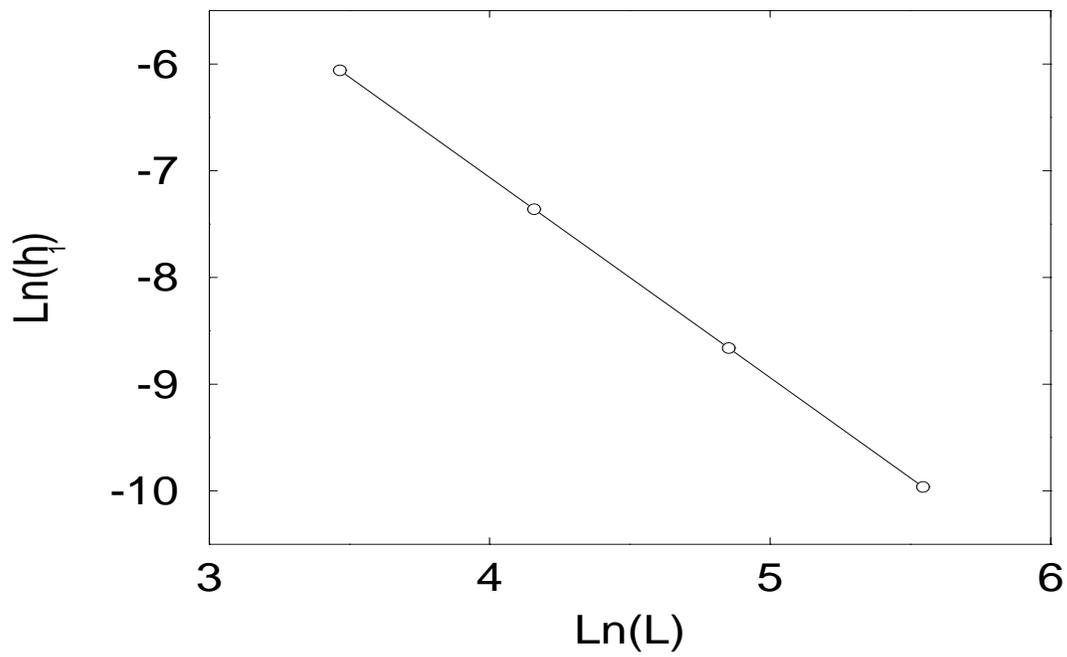}
\caption{Leading FSS of the First Lee--Yang Zeroes $h_1(L)$ 
for the $XY$--model at $\beta = 1.11$.}
\label{fig02:xy:lambda_at_1.11}
\end{figure}

\begin{figure}[htb]
\vspace{13cm}
\includegraphics{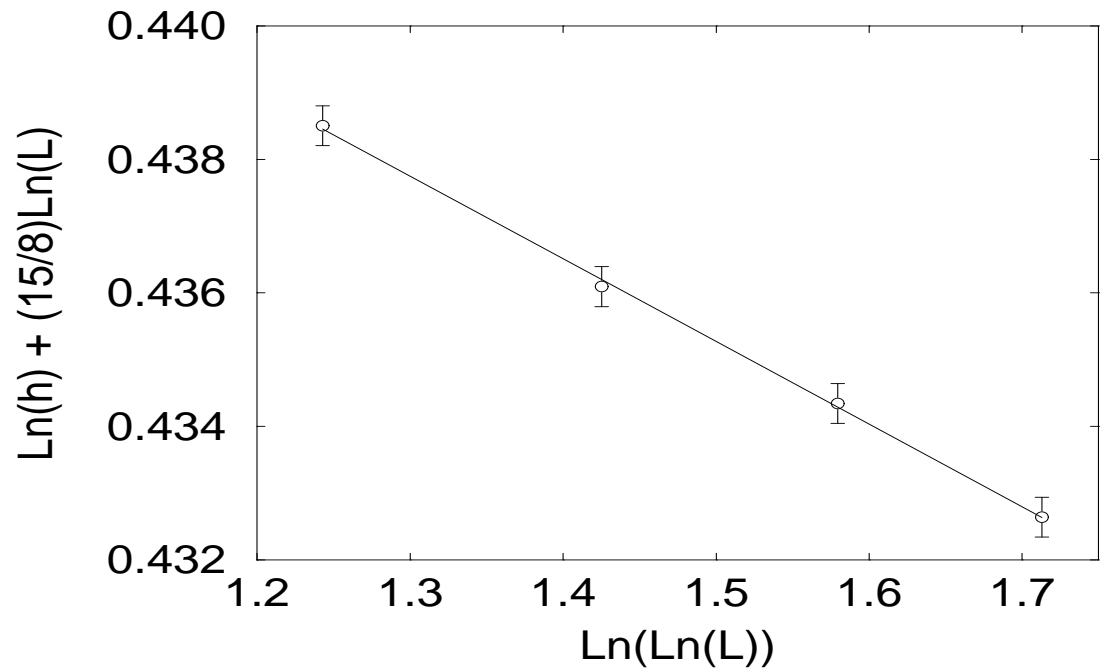}
\caption{Corrections to FSS of the First Lee--Yang Zeroes
for the $XY$--model  at $\beta = 1.11$.}
\label{fig03:xy:r_at_1.11}
\end{figure}

We begin our analysis with the first Lee--Yang zeroes at the 
preliminary critical coupling estimate $\beta_{\rm{RW}} \simeq 1.11$. 
In Fig.~\ref{fig02:xy:lambda_at_1.11} we plot the logarithm 
of the position of the first Lee--Yang zero against the logarithm 
of the lattice size $L$, at $\beta=1.11$.
In the absence of any corrections, the slope 
should give the leading power--law  exponent $\lambda$.    
In fact at $\beta=1.11$ the slope is $\lambda_{\rm{eff}} = -1.8778(2)$.  
The deviation from the KT value of $-15/8 = -1.875$ can be 
attributed to (a) the presence of logarithmic corrections
and/or (b) the possibility that $\beta=1.11$ is not in fact the
critical point. As a preliminary investigation of (a)
we identify the correction exponent $r$ in  (\ref{testr}),  
by plotting $\ln{(\theta_1 L^{15/8})}$  
against  $\ln{(\ln{L})}$   in    Fig.~\ref{fig03:xy:r_at_1.11}. A 
straight line is identified.   Its  slope  is $-0.012(1)$ 
giving evidence for a non--zero value of 
$r$,   albeit   not  in   agreement  with the renormalization 
group predictions of 
$-1/16 = -0.0625$ from \cite{KT}.

The above estimate $\beta_{\rm{RW}}  \simeq 1.11$ for the critical 
coupling came from an analysis in which the possibility of the 
existence of logarithmic corrections was ignored. At this point, 
we must allow for the possibility of the true critical coupling
$\beta_c$ being different than $\beta_{\rm{RW}}$. To systematically 
investigate the extent to which the above conclusions  depend on the 
measurement of $\beta_c$, we now study the above FSS (fits for 
$\lambda_{\rm{eff}}$) and corrections to FSS 
(fits for $r$) as a function of the assumed critical beta.

\begin{figure}[htb]
\vspace{13cm}
\includegraphics{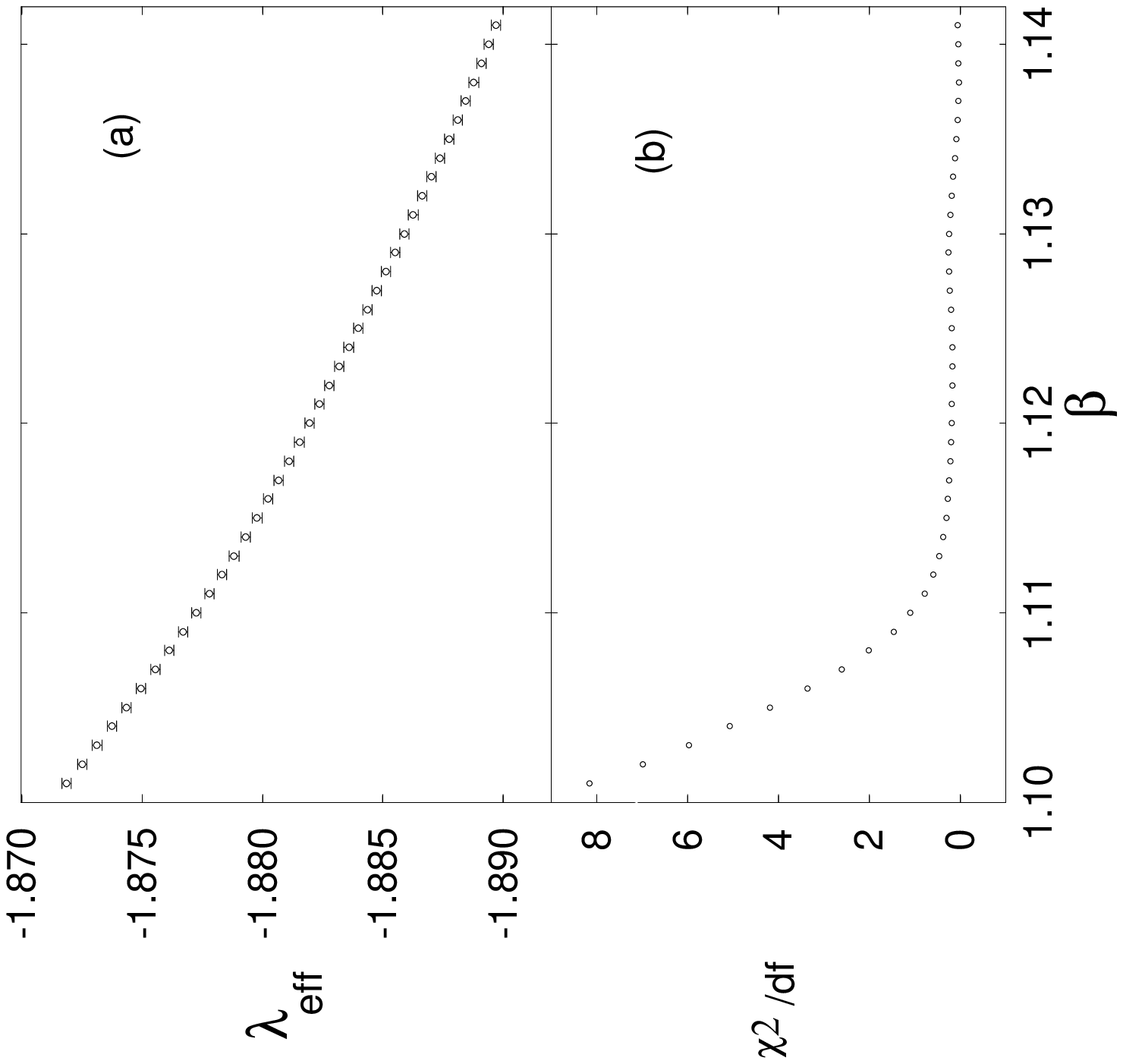}
\caption{Test of the Expected Leading Critical Behaviour
for the $XY$--model: (a) The effective 
exponent $\lambda_{\rm{eff}}$ (slope of a straight line fit) and (b) 
the
corresponding $\chi^2 / {\rm{dof}}$ versus $\beta$.}
\label{fig04:xy:lambda}
\end{figure}

In Fig. 4a we plot $\lambda_{\rm{eff}}$ coming from fits to the 
leading FSS behaviour as a function of $\beta$. Fig. 4b gives 
the corresponding  $\chi^2$ per degree of freedom 
[$\chi^2/{\rm{dof}}$] for these fits.  Not all fits
yield an acceptable $\chi^2/{\rm{dof}}$. To be conservative, 
we take  \lq acceptable\rq{} to mean 
\begin{equation}
\chi^2 / {\rm{dof}} 
\quad
{\rm{\raisebox{-.75ex}{ {\small \shortstack{$<$ \\ $\sim$}} }}}
\quad 
2
\quad ,
\label{criterion}
\end{equation}
which corresponds to a minimum 
confidence level of 14\%. For Fig.~\ref{fig04:xy:lambda} we have  
acceptable   values of $\chi^2/{\rm{dof}}$ only for 
$\beta$ \raisebox{-.75ex}{ {\small \shortstack{$>$ \\ $\sim$}} } 
$1.107$.
The figure indicates that FSS sets in above this value. I.e., the 
system remains critical for 
$\beta$ \raisebox{-.75ex}{ {\small \shortstack{$>$ \\ $\sim$}} } 
$1.107$.

This range of critical $\beta$ values corresponds to
$\lambda_{\rm{eff}}$ \raisebox{-.75ex}{ {\small \shortstack{$<$ \\ $\sim$}} }
$-1.8761(2)$. This range of critical $\beta$ values 
does not include that  for which the KT value of 
$-15/8 = -1.875$ is recovered (Fig.~\ref{fig04:xy:lambda} gives 
$\lambda_{\rm{eff}} \simeq -1.875$ at $\beta \simeq 1.105$ with 
$\chi^2/{\rm{dof}} \simeq 3.4$). Thus we can conclude that ignoring 
logarithmic corrections
gives deviation from the KT value of $-15/8 = -1.875$ for $\lambda$.

\begin{figure}[htb]  
\vspace{13cm}
\includegraphics{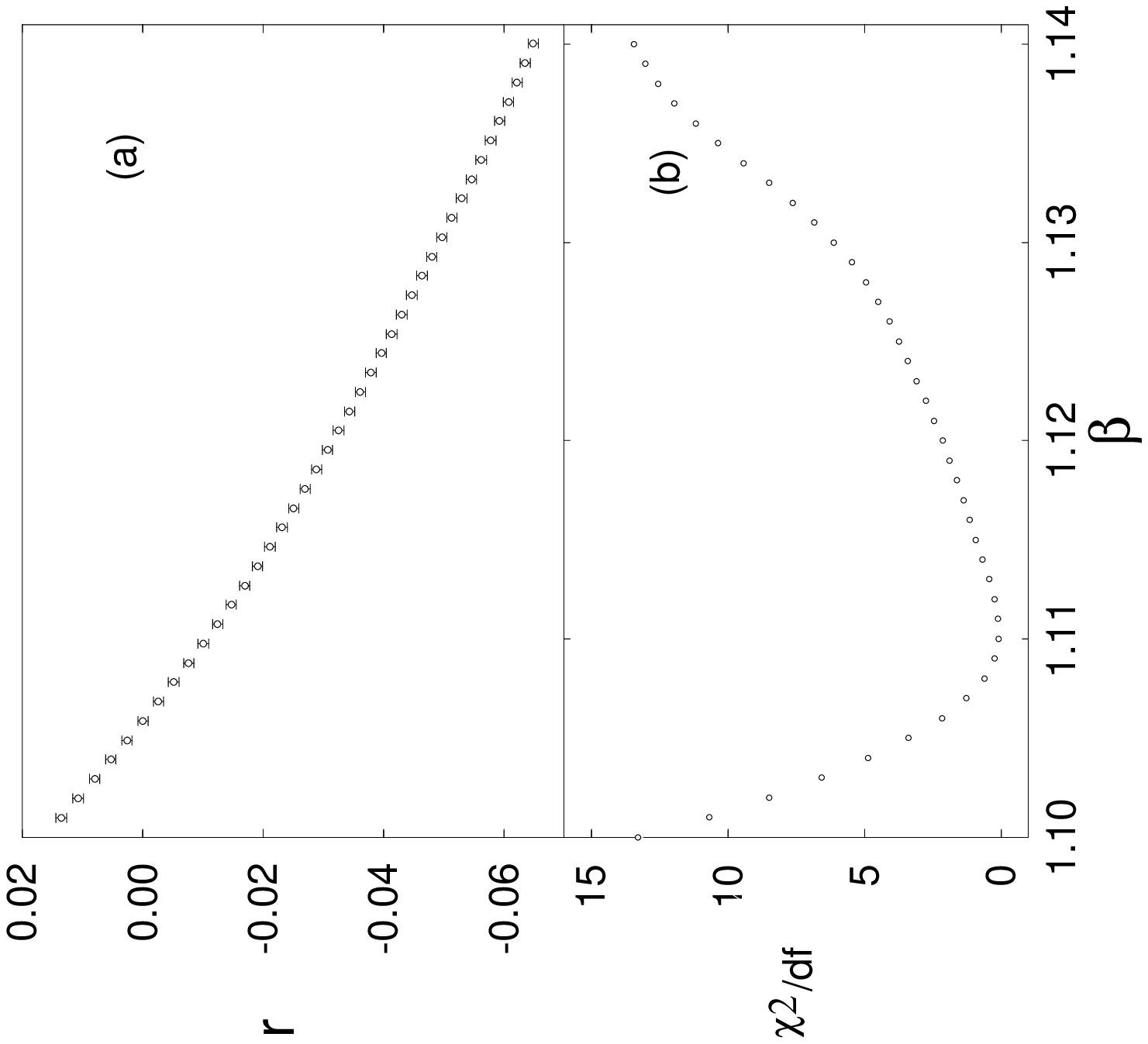}
\caption{Logarithmic corrections
for the $XY$--model: (a) the logarithmic correction
exponent $r$ to the Yang--Lee edge is shown as a function of 
the assumed critical coupling $\beta_c$ and (b) the corresponding 
$\chi^2/{\rm{dof}}$.}
\label{fig05:xy:r}
\end{figure}

Next we investigate how our conclusions regarding the correction 
exponent $r$ depend on the critical coupling. Fig.~\ref{fig05:xy:r} 
gives the results from fits to $r$
from (\ref{testr}) and the corresponding $\chi^2/{\rm{dof}}$. 
we see that acceptable values
are possible only for 
$1.107$ \raisebox{-.75ex}{ {\small \shortstack{$<$ \\ $\sim$}} } 
$\beta_c$
\raisebox{-.75ex}{ {\small \shortstack{$<$ \\ $\sim$}} } $1.119$.  
The $r=0$ solution again corresponds to $\beta_c \simeq 1.105$ and
$\chi^2/{\rm{dof}} \simeq 3.4 $. Similary, we find that a fit with 
$r=-1/16$ would correspond to $\beta_c \simeq 1.138$ and 
$\chi^2/{\rm{dof}} \simeq 13$.
 
Therefore we have numerical evidence challenging the detailed 
quantitative predictions of Kosteritz and Thouless.
In summary, the results of the analysis of the first 
Lee--Yang zeroes of the $XY$--model is that
assuming the KT value $\eta=1/4$, we find non-zero 
logarithmic corrections to scaling and a corresponding
estimate of the critical temperature:
\beq
r=-0.02(1) \, ,\qquad \beta_c=1.113(6)\, . 
\label{eq:results}
\eeq

\paragraph{KT Versus Power Law Scaling Scenario}
The KT essential scaling scenario was called into question in 
\cite{SeSt88} where numerical evidence in support of  conventional 
power law scaling behaviour in the $Z(10)$ clock model was given. 
While this model is expected to be in the same universality class 
at the $XY$--model (same critical indices) it may have a different 
critical coupling $\beta_c$ hindering a comparison with  our numerical 
results. However a recent paper by Kim \cite{Kim} also  claims to 
favour 
a convential as opposed to KT phase transition in the $XY$--model.

It is extremely difficult to distinguish between KT and power law 
scaling on the basis of numerical methods alone \cite{EdGo91}. 
Power law scaling for the thermodynamic functions (with additive 
power law corrections) would lead to (\ref{nojosephson}) being 
replaced by Josephson's law
\begin{equation}
  \alpha = 2 - \nu d
\quad ,
\label{Josephson}
\end{equation}
and the FSS of the zeroes (\ref{testr}) by
\begin{equation}
  \theta_1(\beta_c) = a_1 L^\lambda \left( 1 + a_2 L^{a_3}\right)
\quad ,
\label{PP}
\end{equation}
for some $a_1$, $a_2$ and $a_3$.
If $\nu > 1$ the singular part of the specific heat decreases with
lattice size as in the KT case (\ref{cwlog}). Thus Josephson's law 
may not be a numerically  useful criterion for distinguishing  
between  the two scaling scenarios. 

The leading FSS of $\theta_1$ in (\ref{PP}) is the same as for
the KT case (\ref{testr}) and  ignoring corrections leads to 
the same conclusions regarding $\lambda_{\rm{eff}}$. However, as 
seen above, it is essential to include corrections to correctly 
identify $\beta_c$. Ignoring corrections would lead us to 
conclusions similar to \cite{Kim}. The numerical analysis 
of \cite{Kim} applies to $\xi$ and $\chi$.  There, 
$\eta (\beta)$ is measured by fitting to the FSS formula
$\chi_L \sim L^{2 - \eta (\beta)}$ (ignoring multiplicative 
logarithmic corrections). Good fits ($\chi^2/{\rm{dof}} < 0.5$) 
were reported for  $\beta > 1.105 $. Using
\begin{equation}
 \lambda = -2 + \eta / 2 
\quad ,
\label{etalambda}
\end{equation}
the corresponding $\lambda_{\rm{eff}}$  can be found. These are 
listed and compared to our own results (when corrections are 
ignored) in Table~\ref{tabb}.
The agreement is impressive. 

\noindent
\begin{table}[ht]
\caption{
    Comparison of our results for 
    $\lambda_{\rm{eff}}$ 
    (ignoring multiplicative logarithmic corrections) 
    to  [27].
}
\label{tabb}
\begin{center}
\vspace{0.5cm}  \noindent\begin{tabular}{|l|l|l|l|l|l|} 
    \hline 
    \hline
 $T$ & $\beta = 1/T$  & $\eta$ from \cite{Kim} & 
 $\lambda_{\rm{eff}}$ from \cite{Kim}  
 & Our $\lambda_{\rm{eff}}$  & Our $\chi^2/{\rm{dof}}$ \\
      & & & & &  \\
    \hline
    0.910 & 1.099 & 0.2600(10) & -1.8700(10) & -1.8712(2) & 9.0
 \\
    \hline
    0.905 & 1.105 & 0.2520(05) & -1.8740(05) & -1.8750(2) & 3.4            
 \\
    \hline
    0.900 & 1.111 & 0.2450(05) & -1.8770(05) & -1.8778(2) & 0.8            
 \\
    \hline
    0.890 & 1.1236 & 0.2310(10) & -1.8840(10) & -1.8838(2) & 0.2            
 \\
    \hline
    \hline
  \end{tabular}
\end{center}
\end{table}

In \cite{Kim} the critical point is located by finding a coupling,
$\beta_{\rm{P}}$, where $\chi_L \sim L^{2 - \eta (\beta)}$  holds 
for $\eta = 1/4$. The result is $1.105 < \beta_{\rm{P}} <1.111$ 
with `most probably'  $\beta_{\rm{P}}= 1.1062(6)$
($T_c = 0.9040(5)$).  This compares well to our result
$r=0$ (no corrections) at $\beta \simeq 1.105$. Using  this 
as the critical coupling, \cite{Kim} reports that fits to $\xi$ 
in (\ref{ktxi}) give an unacceptable $\chi^2/{\rm{dof}}$.
In the light of our analysis this is hardly surprising. The system
is still in fact in the high temperature
phase and criticality has not been reached -- $\beta_{\rm{P}}$
is not the true critical coupling.

Examination of power law scaling in \cite{Kim} in the form 
of a four parameter fit to $\xi_\infty = At^{-\nu}(1 + bt^w)$
yields  a reasonable $\chi^2/{\rm{dof}}$. We attempted a three 
parameter fit to (\ref{PP}) and found acceptable 
$\chi^2/{\rm{dof}}$ over a range of values for $\beta$ but 
believe this to be attributable to the fact that it is easy to 
make  a three parameter fit to only 4 data points. 
So while we cannot logically exclude a power law scaling scenario
on the basis of our numerical results neither can the (modified) 
KT scaling scenario be excluded on the basis of \cite{Kim}. 

The lesson here is that it is essential to include consideration
of logarithmic corrections to KT scaling in order  not to be led 
to the wrong value of $\beta_c$ upon which conclusions are highly 
sensitive.

\subsubsection{The Step Model}

\paragraph{Scaling of Lowest Lee--Yang Zeroes}
The analysis began with a rough search for the leading critical 
behaviour predicted by (\ref{testr}). An independent test was also 
made using the (less accurate) susceptibility data and  
(\ref{chifss}). Both methods indicated critical behaviour setting in 
for $\beta$  
\raisebox{-.75ex}{ {\small \shortstack{$>$ \\ $\sim$}} } $1.2$. 

In Fig.~6 we display the results for the effective exponent 
$\lambda_{\hbox{eff}}^{\rm{step}}$ as a function of 
$\beta$ together with the corresponding $\chi^2/{\rm{dof}}$.
Acceptable fits are only found for 
$\beta_c^{\rm{step}} 
\raisebox{-.75ex}{ {\small \shortstack{$>$ \\ $\sim$}} } 1.185$.
We note that the corresponding  values of 
$\lambda_{\hbox{eff}}^{\rm{step}}$ 
($ \raisebox{-.75ex}{ {\small \shortstack{$<$ \\ $\sim$}} } 
-1.8715(2)$) 
include that ($-15/8 = -1.875$) corresponding to the KT prediction.

Again, the evidence for critical behaviour is the existence of a 
range of acceptable chi-squared values for a linear fit. In view 
of the similarity of $\lambda_{\hbox{eff}}^{\rm{step}}$ to the 
expected KT value, we now proceed to test  the further hypothesis 
that the step model is in fact in the same  universality class as 
the $XY$--model. We {\em assume} the behaviour 
(\ref{testr})  with $\lambda^{\rm{step}} = -15/8$
and search for corrections in Fig.~\ref{fig09:step:r}.
We find acceptable fits for
$ 1.195\leq\beta_c^{\rm{step}}\leq 1.295$
with $0.009 \geq r \geq -0.034$.

The range of acceptable $r$ values includes that found earlier for 
the $XY$--model and that corresponding to no logarithmic corrections 
($r=0$). Again, this range excludes the  prediction $r = -1/16$ 
coming from the approximate renormalisation group treatment of 
the $XY$--model \cite{KT}. Thus we conclude that the present data 
are compatible with the step model being in the same universality 
class as the $XY$--model. We do not, however,
exclude other possibilities.

In summary,
\begin{eqnarray}
 XY{\rm{-Model:}}   &   
 1.107 \raisebox{-.75ex}{ {\small \shortstack{$<$ \\ $\sim$}} } 
 \beta_c \raisebox{-.75ex}{ {\small \shortstack{$<$ \\ $\sim$}} }
 1.119 \quad , &
 -0.01 \raisebox{-.75ex}{ {\small \shortstack{$>$ \\ $\sim$}} }
 r \raisebox{-.75ex}{ {\small \shortstack{$>$ \\ $\sim$}} }
 -0.03
\quad ,
\\
{\rm{Step ~ Model:}}   &
 1.195 \raisebox{-.75ex}{ {\small \shortstack{$<$ \\ $\sim$}} }
 \beta_c^{\rm{step}} 
\raisebox{-.75ex}{ {\small \shortstack{$<$ \\ $\sim$}} }
 1.295 \, , &
 0.01 \raisebox{-.75ex}{ {\small \shortstack{$>$ \\ $\sim$}} }
 r^{\rm{step}} 
\raisebox{-.75ex}{ {\small \shortstack{$>$ \\ $\sim$}} }
 -0.03
\, .
\end{eqnarray}

\subsubsection{Higher Order Corrections} 

We have shown that a FSS analysis ignoring multiplicative 
logarithmic corrections leads to deviation from the KT 
prediction $-15/8$ for $\lambda$ or, equivalently, $1/4$ 
for $\eta(\beta_c)$. When multiplicative logarithmic corrections 
are taken into account the results are compatible with 
$\eta(\beta_c) = 1/4$ but, now, the leading logarithm exponent $r$
is not that given by KT.

In the same spirit, it is conceivable that numerical measurement 
of these leading logarithmic 
corrections may become compatible with the KT RG predictions 
($r=-1/16$) if higher order corrections are accounted for.

Keeping higher order (additive) corrections in the RG treatment of 
the magnetic susceptibility modifies the right hand side of 
(\ref{KTchi7}) by \cite{AmGo80,KaZi81}
\begin{equation}
 \left\{ 1 + 
O \left( \frac{\ln{\ln{\xi_\infty}}}{\ln{\xi_\infty}} \right)\right\}
\quad .
\label{loglogcorrectionschi}
\end{equation}
The FSS behaviour of the Yang--Lee edge is now expected to be (from 
(\ref{as2}))
\begin{equation}
 \theta_1(\beta_c) = a_1 L^\lambda (\ln{L})^r
 \left\{ 1 + a_2 \frac{\ln{\ln{L}}}{\ln{L}} \right\}
\quad .
\label{loglogcorrectionsedge}
\end{equation}
Unbiased four parameter fits are hampered by the convergence 
problems mentioned in section 1.1.
However, if we again accept the leading scaling behaviour                      
$\lambda = -15/8$, two parameter fits to $a_1$ and $a_2$ 
in (\ref{loglogcorrectionsedge}) yield
unacceptable $\chi^2/{\rm{dof}}$ if $r$ is $-1/16$ in both the
$XY$-- and step models. In fact, $\chi^2/{\rm{dof}}$ ranges
from $13$ to $2$ as $\beta$ decreases from $1.14$ to $1.10$
in the $XY$--model and from $9$ to $2.5$  as $\beta$ decreases 
from $1.40$ to $1.17$ in the step model, indicating that such fits
may become acceptable for very low $\beta$ in both cases.
(Note that in the case of the $XY$--model such low $\beta$ 
is outside the range of $\beta_c$--measurements in table 1).
On the other hand, if we fix
$r=-0.02$ we find $\chi^2 / {\rm{dof}}
{\rm{\raisebox{-.75ex}{ {\small \shortstack{$<$ \\ $\sim$}} }}}  2$      
over the whole $\beta$--range in both models.

%

\section{Analysis of Higher Index Zeroes}

In the calculation leading to (\ref{as2}) we assumed that the low
index zeroes have the same FSS behaviour. We now wish
to (a) check that assumption, (b) redo the whole analysis for
each index $j=2,\dots,10$ separately and (c) introduce a new 
concept we call `index scaling' and apply it to 
these first ten zeroes. 
In fact we can determine up to fifteen zeroes for each lattice 
size. However the errors for the 
highest index zeroes become too large to allow reliable analysis.

\subsection{Analyses of Individual Higher Index Zeroes}
\label{4.2}

\begin{table}[ht]
\caption{The results of the $\lambda_{\rm{eff}}$--analysis for the 
$j$--index zeroes for the $XY$-- and step models: 
The criterion (\ref{criterion}) holds for each $j$ for
$\beta$ above and $\lambda$ below the values listed.}\label{tabc}
\vspace{0.5cm}
\begin{center}
\noindent\begin{tabular}{|c|c|c|c|c|}
    \hline
    \hline
  & \multicolumn{2}{c|}{}
             & \multicolumn{2}{c|}{} \\
  & \multicolumn{2}{c|}{$XY$--Model} 
             & \multicolumn{2}{c|}{step Model} \\
    \hline
 $j$ & $ \beta $ & $ \lambda_{\rm{eff}} $ 
            & $ \beta $ & $ \lambda_{\rm{eff}} $     \\
 \hline
01  &   1.107  &  -1.8761(2)  &   1.185  &  -1.8715(2)
\cr
02  &   1.107  &  -1.8761(2)  &   1.185  &  -1.8716(2)
\cr
03  &   1.106  &  -1.8754(2)  &   1.180  &  -1.8716(2)
\cr
04  &   1.106  &  -1.8753(2) &   1.185  &  -1.8717(2) 
\cr
05  &   1.104  &  -1.8741(2) &   1.190  &  -1.8725(2)
\cr
06  &   1.103  &  -1.8735(2) &   1.200  &  -1.8739(2)
\cr
07  &   1.103  &  -1.8735(2) &   1.215  &  -1.8757(2)
\cr
08  &   1.103  &  -1.8735(2)  &   1.215  &  -1.8757(2)
\cr
09  &   1.100  &  -1.8717(2)  &   1.195  &  -1.8735(2)
\cr
10  &   1.100  &  -1.8717(2) &   1.185  &  -1.8722(2)
\cr
    \hline
    \hline   
\end{tabular}
\end{center}
\end{table}

\small
\begin{table}[ht]
\caption{The Results of the $r$--Analysis for the $j$--Index
Zeroes for the $XY$-- and step models: 
The criterion (\ref{criterion}) holds for each $j$ for the 
following $\beta$ ranges and corresponding $r$ range.}\label{tabd}
\vspace{0.5cm}
\begin{center}
\noindent\begin{tabular}{|c|c|c|c|c|}
    \hline
    \hline
  & \multicolumn{2}{c|}{}
             & \multicolumn{2}{c|}{} \\
  & \multicolumn{2}{c|}{$XY$--Model}
             & \multicolumn{2}{c|}{step Model} \\
    \hline
 $j$ & $ \beta $ &  $r$ & $ \beta $ &  $r$ \\
 \hline
01  &   1.107--1.119 &  -0.005(1)   --   -0.030(1)
    &   1.195--1.295 &   0.009(1)   --   -0.034(1)
\cr
02  &   1.106--1.120 &  -0.002(1)   --   -0.032(1)
    &   1.195--1.295 &   0.009(1)   --   -0.034(1)
\cr
03  &   1.106--1.120 &  -0.002(1)   --   -0.031(1)
    &   1.195--1.295 &   0.008(1)   --   -0.034(1)
\cr
04  &   1.105--1.119 &   0.001(1)   --   -0.029(1)
    &   1.195--1.295 &   0.008(1)   --   -0.034(1)
\cr
05  &   1.104--1.117 &   0.004(1)   --   -0.024(1)
    &   1.200--1.295 &   0.005(1)   --   -0.034(1)
\cr
06  &   1.104--1.116 &   0.004(1)   --   -0.022(1)
    &   1.205--1.295 &   0.002(1)   --   -0.034(1)
\cr
07  &   1.104--1.116 &   0.004(1)   --   -0.022(1)
    &   1.210--1.300 &  -0.001(1)   --   -0.035(1)
\cr
08  &   1.103--1.116 &   0.007(1)   --   -0.021(1)
    &   1.210--1.300 &   0.001(1)   --   -0.035(1)
\cr
09  &   1.102--1.119 &   0.009(1)  --   -0.026(1)
    &   1.200--1.310 &   0.004(1)  --   -0.038(1)
\cr
10  &   1.100--1.122 &   0.014(1)  --   -0.031(1)
    &   1.190--1.350 &   0.009(1)  --   -0.047(1)
\cr
    \hline
    \hline
\end{tabular} 
\end{center}
\end{table}
\normalsize    
    
We repeat the analysis of (\ref{testr}) for each index separately
for $j = 2, \dots , 10$ and for both the $XY$-- and step models.
The results of the analyses for $\lambda_{\rm{eff}}$  are sumarized 
in Table~\ref{tabc}. For the $XY$--model the analysis applied to 
individual higher index zeroes 
($j$ \raisebox{-.75ex}{ {\small \shortstack{$>$ \\ $\sim$}} } $4$)
appears to accomodate the $RG$ result $\lambda = -1.875$. However
higher index zeroes are less accurately determined than lower index 
ones and the $j=1$ results remain the most accurate.

Accepting $\lambda = -1.875$ and looking for corrections from 
applications of the $r$--analysis to each index $j$ separately 
gives the results 
summarized in Table~\ref{tabd}. Acceptable fits are obtained only
in the $\beta$ range given and lead to the listed ranges for $r$.
These $\beta$ and $r$ ranges are graphically represented in 
Fig.~\ref{fig10:xy:j} (for the $XY$--model) 
and Fig.~9 (for the step model).
Although in both models the higher index zeroes individually allow 
for $r=0$,
at no stage (up to $j=10$) do we find results compatable with the 
RG result $-1/16 = -0.0625$. 
So the higher index zeroes analysed individually yield  the same
information with decreasing accuracy as $j$ increases
providing further strong evidence of the validity of our earlier 
conclusions.

Finally, analyses of higher order (additive) corrections for
the individual higher index zeroes yield the same qualitative 
results as those presented in sec. 3.3.4.

\subsection{Analysis of the Density of Zeroes (Index Scaling)}

The last section demonstrates the success of FSS applied to each 
zero separately. We can now be confident that the lowest lying zeroes
scale in the same way with lattice size $L$ and that this scaling 
behaviour is 
\begin{equation}
 \theta_j (\beta_c) \sim L^{\lambda} (\ln{L})^r
\quad .
\label{okay}
\end{equation}
In this paragraph we wish to report an attempt to combine the scaling 
behaviour of the various indexed zeroes.

The density of zeroes for the finite (binned) system is given 
in (\ref{gG}) as a sum over delta functions,
\begin{equation}
 g_L(\beta,\theta) 
 = 
 L^{-d}
 \sum_{j=1}^N \delta(\theta - \theta_j(\beta,L))
\quad ,
\label{I}
\end{equation}
where $N$ is the total number of zeroes. The corresponding 
cumulative density function $G$ is a step function with
\begin{equation}
 G_L(\beta,\theta) = \frac{j}{N}
 \quad \quad \quad {\rm{if}} \quad 
 \theta \in (\theta_j,\theta_{j+1})
\quad .
\end{equation}
This function can be smeared by replacing the delta functions
of (\ref{I}) by Gaussians
\begin{equation}
 g_L(\beta,\theta) 
 = 
 \frac{1}{N}
 \sum_{j=1}^N{
   \frac{1}{\sqrt{2 \pi} \sigma_j(L)}
   e^{
      - \frac{1}{2}
      \left(
            \frac{\theta - \theta_j(\beta,L)}{\sigma_j(L)}
      \right)^2
      } 
}
\quad ,
\end{equation}
where $\sigma_j(L)$ is some appropriate quantity giving the spread 
of the Gaussian around the $j^{\rm{th}}$ zero. The cumulative 
density function is then a sum of error functions
\begin{equation}
 G_L(\beta,\theta) = \frac{1}{2N}
 \sum_{j=1}^N{
             \left(
                     1+ {\rm{erf}}{
                             \left(
                   \frac{\theta - \theta_j}{\sqrt{2} \sigma_j}
                             \right)
                            }
             \right)
            }
\quad .
\end{equation}
If $\sigma_j(L)$ is small enough then
\begin{equation}
 G_L(\beta,\theta_j) = \frac{2j-1}{2N}
\quad .
\label{III}
\end{equation}
In the thermodynamic limit (\ref{twoPhi}) gives 
\begin{equation}
 G_\infty (\beta,\theta)
 \sim
 \xi_\infty^{-d}
 \Phi
 \left(
        \frac{\theta}{\theta_c(\beta)}
 \right) 
\quad .
\label{IV}
\end{equation}
The corresponding FSS formula is 
\begin{equation}
 G_L (\beta_c,\theta) 
\propto
 L^{-d}
 \Phi
 \left(
        \frac{\theta}{\theta_1(\beta_c)}
 \right)
\quad .
\label{IVfss}
\end{equation}
Comparing with (\ref{III}) and using $N \propto L^d$,
\begin{equation}
 \frac{\theta_j}{\theta_1}
 =
 \Phi^{-1}
 \left(
   \Phi (1)
   (2j-1)
 \right)
\quad .   
\label{V}
\end{equation}
The density of zeroes in the thermodynamic limit is
\begin{equation}
 g_\infty (\beta,\theta)
 =
 \frac{dG_\infty (\beta,\theta)}{d\theta}
 =
 \xi_\infty^{-d} \theta_c^{-1}
 \Phi^\prime
 \left(   
        \frac{\theta}{\theta_c(\beta)}
 \right)
\quad .
\end{equation}
In a numerical high temperature and high field study, Kortman and 
Griffiths showed that, for Ising models in the high temperature 
phase, this density exhibits power law behaviour as a function 
of the distance  of $\theta$ away from the edge \cite{KoGr71},
\begin{equation}
 \Phi^\prime
 \left(
        \frac{\theta}{\theta_c(\beta)}
 \right)
 \sim
 \left|
      1  -   \frac{\theta}{\theta_c(\beta)}
 \right|^\sigma
\quad .
\label{IVdd}
\end{equation}
At $\beta = \beta_c$, $\sigma = 1 / \delta$ where  $\delta$ is 
the usual magnetization index. This index governs the response 
of the magnetization to the presence of a weak external magnetic 
field,
\begin{equation}
 M \sim h^{\frac{1}{\delta}} 
\quad .
\label{VI}
\end{equation}
One therefore is led to expect
\begin{equation}
 \Phi (x) \sim (1-x)^{\frac{\delta + 1}{\delta}}
\quad , 
\end{equation}
or,
\begin{equation}
 \Phi^{-1} (y) - 1 \sim y^{\frac{\delta}{\delta + 1}}
\quad .
\end{equation}
There are additive higher order corrections to formulae 
(\ref{IV})--(\ref{VI}). 
Now, (\ref{V}) gives
\begin{equation}
 \frac{\theta_j}{\theta_1}
 =
 c_1
 + 
 c_2 
 (2j-1)^{\frac{\delta}{\delta + 1}}
\quad ,
\label{indexd}
\end{equation}
for some $c_1$ and $c_2$ which incorporate the higher order 
corrections.  Using the standard scaling relation
\begin{equation}
 \delta = \frac{d+2-\eta}{d-2 + \eta}
\quad ,
\end{equation}
we can express the leading behaviour in terms of $\lambda$
\begin{equation}
 \frac{\theta_j}{\theta_1}
 =
 c_1
 +
 c_2
 (2j-1)^{-\frac{\lambda}{d}}
\quad .
\label{index}
\end{equation}
Again, one expects there to be additive volume dependent corrections 
to this formula. 

Nonetheless, to leading order the $L$--dependence drops out of the 
ratio of the $j^{\rm{th}}$ to  $1^{\rm{st}}$ zeroes. This opens up 
the possibility of a new type of scaling technique -- the use of 
the dependency of this ratio on the index $j$ to find critical 
indices. If successful, this method could be applied in parallel 
to traditional FSS.  The advantage would be that one doesn't need 
enormous lattices to eliminate  finite--size effects, nor does one 
need a large range of lattice sizes to  exploit FSS. In cases where
only one lattice size is available, index scaling may, in principle,
still yield a result.

We can check the form (\ref{indexd}) in one dimension.
From the exact solution to the one dimensional Ising model with
periodic boundary conditions, the Lee--Yang zeroes are given by 
\cite{LY}
\begin{equation}
 \cos{(2 \theta_j)}
 =
 -u^2
 +
 (1-u^2)
 \cos{\left( \frac{2j-1}{L} \pi \right)}
\quad ,
\end{equation}
where $u = \exp{(- 2 \beta)}$. In one dimension the critical point 
is at zero temperature or $u_c = 0$. Then the LY zeroes are at
\begin{equation}
 \theta_j (\beta_c) = \pm \frac{2j-1}{2L} \pi
\quad .
\end{equation}
Then,
\begin{equation}
 \frac{\theta_j}{\theta_1}
 =
 2j - 1
\quad ,
\label{index1d} 
\end{equation}
which is of the form (\ref{indexd}) with $\delta = \infty$,
the correct value for $\delta$ in one dimension.
There is no exact solution to the Ising model in the presence of
an external field in two or more dimensions.

\paragraph{The $XY$--Model}
In Fig.~10 
we test (\ref{index}) for the two dimensional $XY$--model at 
$\beta_c = 1.115$. The plot involves ten LY zeroes for each of 
the four lattice sizes -- forty data points in all. 
As expected, no $L$--dependence is visible. Fitting to 
\begin{equation}
 \frac{\theta_j}{\theta_1}
 =
 c_1 
 +
 c_2
 (2j-1)^{c_3}
\label{indexa}
\end{equation}
gives, however, $c_3 = 0.9827(3)$,
5\%  away from the expected value $-\lambda / d = 0.9375$
and a rather poor  $\chi^2/{\rm{dof}}$ ($8.7$). The 
line in Fig.~\ref{fig11:xy:index} is this best fit
to (\ref{indexa}).

In an early paper on the scaling of partition function zeroes 
\cite{IPZ} it was suggested that the scaling variable is in 
fact $j/L^d$.  When we plot $h_j(L)$ against $j/L^d$ we find 
the data does not collapse onto a smooth curve. A later paper 
showed that this dependency only holds for high index zeroes 
\cite{GlPrSc87}. Indeed our result
\begin{equation}
 \theta_j = \theta_j \left( \frac{2j-1}{L^d} \right)
\label{dependency}
\end{equation}
asymptotically  approaches that of \cite{IPZ} at high $j$.
However, the poor $\chi^2/{\rm{dof}}$ for the fit to 
the form (\ref{indexa})
 indicates that while it might be nearer the truth than
\cite{IPZ}, (\ref{dependency}) is also not the full 
story\footnote{
Other arguments for this 
functional form are contained in \cite{Le94}
}
.

\begin{table}[ht]
\caption{The Results of the index scaling Analysis 
small ranges of $j$.
}\label{tabe}
\vspace{0.5cm}
\begin{center}
\noindent\begin{tabular}{|c|c|}
    \hline
    \hline
 $j$ & $ c_3 $ \\
     & \\
 \hline
2-5  &   0.991(1)
\cr
3-6  &   0.986(1)
\cr
4-7  &   0.981(1)
\cr
5-8  &   0.977(1)
\cr
6-9  &   0.973(1)
\cr
7-10 &   0.969(2)
\cr
    \hline
    \hline
\end{tabular}
\end{center}
\end{table}

To see how $c_3$ depends on the $j$--indices used, we make fits to 
(\ref{indexa}) for $j = 2-5$, $3-6$ and so on to $7-10$. We have a 
three parameter fit to $4 \times 4 = 16$ data points in each case.
The fits are now of good quality ($\chi^2/{\rm{dof} < 1}$) and the
results for $c_3$ are summarised in Table~\ref{tabe}. The 
j--dependence indicates that (\ref{indexa}) can only be true 
asymptotically and encourages further investigation into the 
the full scaling form. A similar picture is found for the step model.

\section{Conclusions}
   
\setcounter{equation}{0}

We have presented a very general (model independent) theoretical 
method to test the self--consistency of the scaling behaviour of 
odd and even thermodynamic functions. Application of this method 
to the KT scaling predictions for the two  dimensional $XY$--model 
reveals that multiplicative logarithmic corrections cannot be ignored. 
In the case of specific heat these logarithmic corrections were 
identified analytically. 
The corrections corresponding to the magnetic susceptibility  were 
identified numerically from the scaling behaviour of the lowest 
Lee--Yang zeroes.


The theoretical and numerical techniques we have developed are 
fully independent of the physics of the phase transition. The 
conventional picture is that the $XY$--model phase transition 
is driven by a vortex binding mechanism \cite{Be71,KT}. This 
is in  contrast to the  customary picture of the step model
\cite{GuJoTh72,GuJo73,LeSh87,LeSh88,NyIr86,SVWi88,GuNy78}. 
Here, while the configuration space is the same as that of 
the $XY$--model, vortex formation is not believed to be 
energetically favourable given the discontinuous nature of the
interaction function. If the $XY$--model phase transition is 
truely driven by a vortex interaction mechanism and if a phase 
transition exists at all in the step model then, barring a
remarkable coincidence, they are expected to belong to different 
universality classes (different scaling  behaviour).

Application of our numerical techniques to the two dimensional 
step model reveals  the critical parameters (including  
logarithmic corrections) to be compatable with those of 
the $XY$--model. This is a strong indicaton  that both
models do in fact belong to the same universality class. 
This conclusion is reinforced by analysis of higher index zeroes.

This result raises questions on how  the vortex binding scenario 
can be the driving
mechanism for the phase transition in the $XY$--model.
Furthermore, the multiplicative
logarithmic corrections identified numerically
by the analysis of up to the first ten Lee--Yang zeroes
are {\em{not}} in agreement with the renormalization group
predictions of Kosterlitz and Thouless.

\vspace{0.5cm}

\paragraph{Note}
Since the preprint version of this paper first appeared,
other authors have reanalysed their data to look for logarithmic 
corrections in the two dimensional $XY$--model. In particular, 
Janke \cite{Ja96} has applied FSS to the Villain version and found 
$r=-0.027(1)$ in agreement with our result. High temperature 
measurements \cite{Ja96} give $r=+0.0560(17)$, close to 
Patrascioiu and Seiler's result $r=+0.077(46)$ \cite{PaSe96}
and $r = +0.042(5)$ to $+0.05(2)$ from Campostrini et al. 
\cite{CaPe96}.

\vspace{1.5cm}

We wish to thank W. Janke, E. Klepfish and C.B. Lang 
for stimulating discussions.

\newpage


\begin{figure}[htb]
\vspace{13cm}
\includegraphics{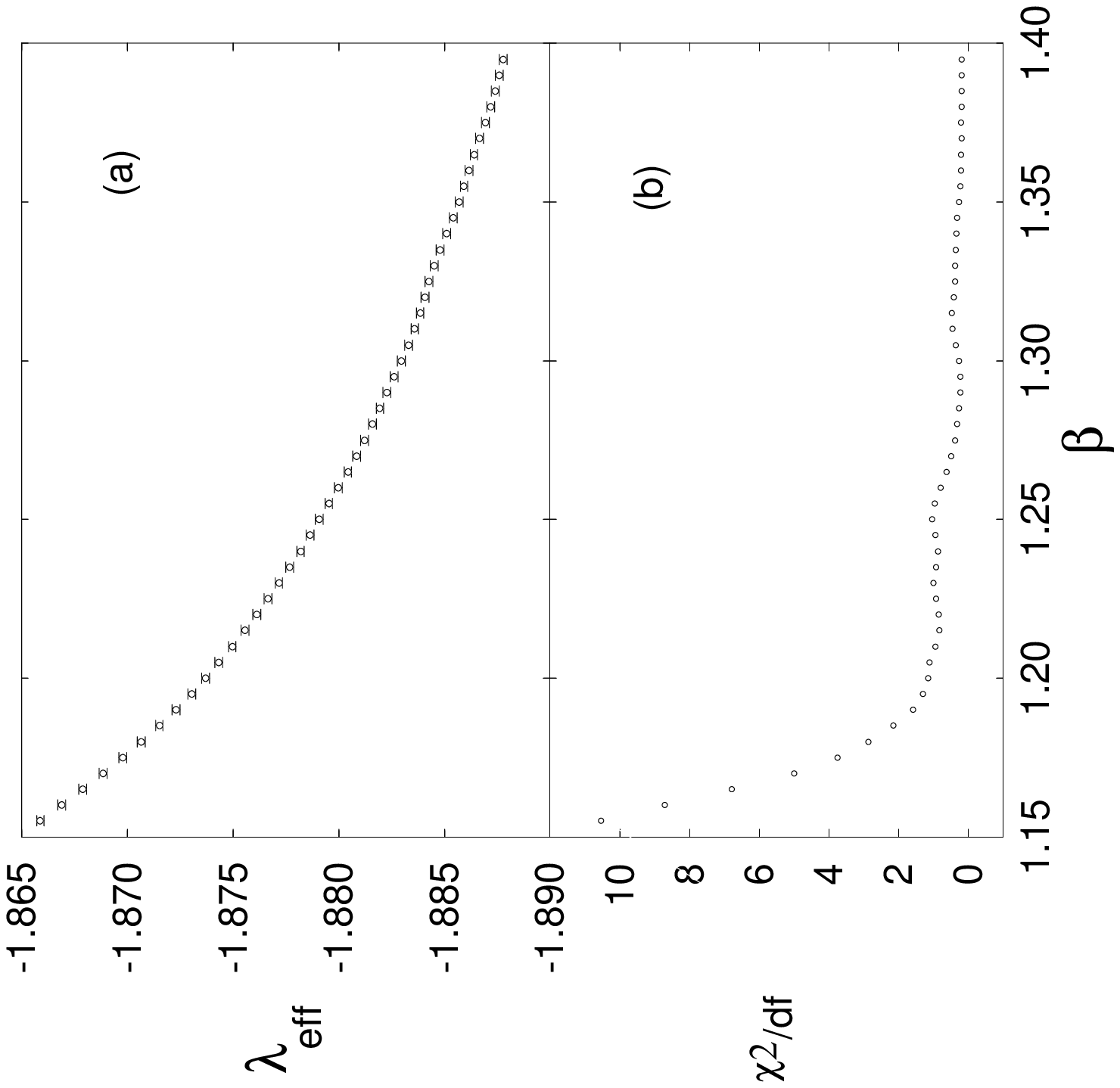}
\caption{Test of the Expected Leading Critical Behaviour
for the step model: (a) The effective 
exponent $\lambda_{\rm{eff}}^{\rm{step}}$ (slope of a 
straight line fit) and (b) the
corresponding $\chi^2 / {\rm{dof}}$ versus $\beta$.}
\label{fig08:step:lambda}
\end{figure}

\begin{figure}[htb]  
\vspace{13cm}
\includegraphics{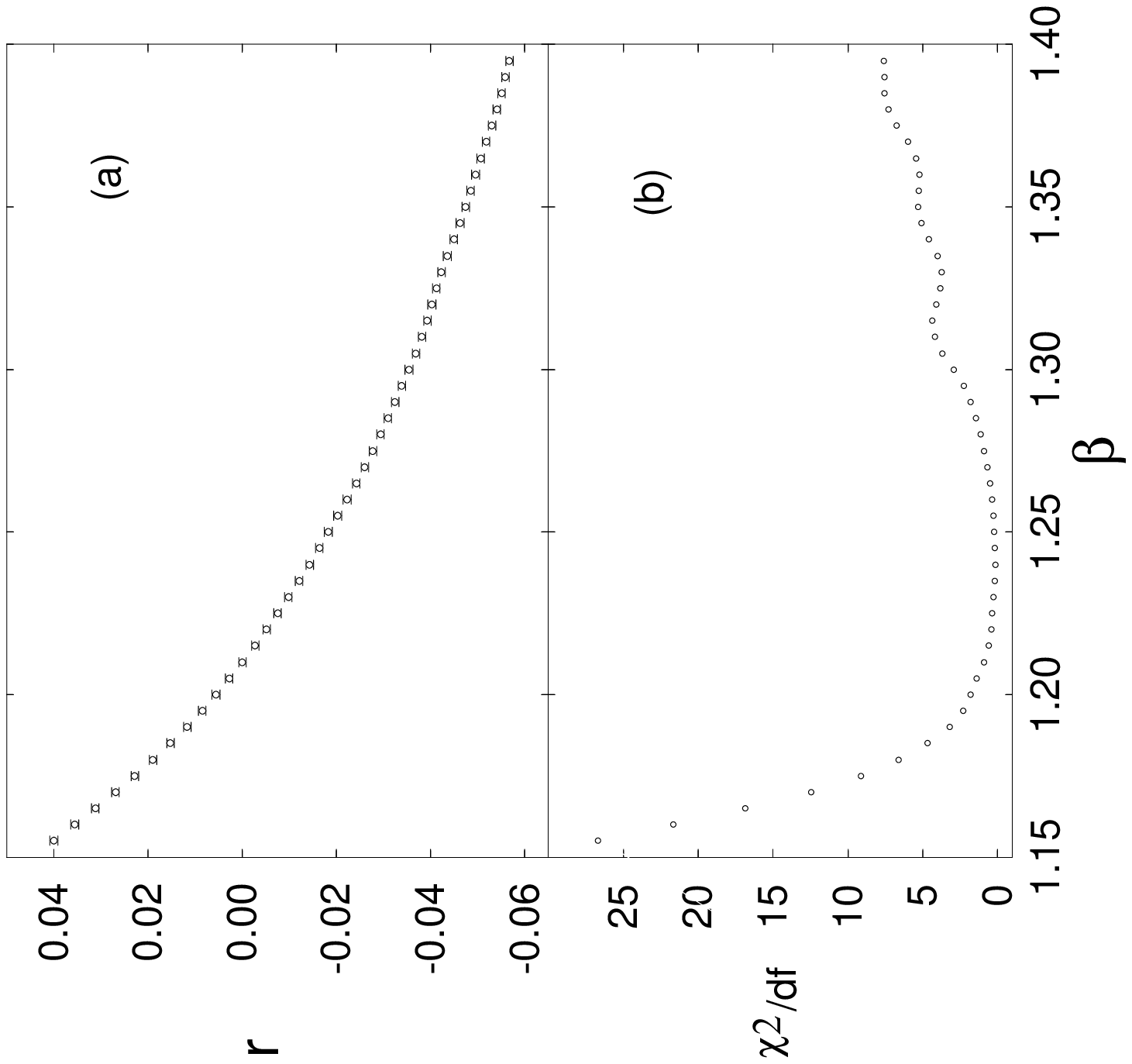}
\caption{Logarithmic corrections for the step model: (a) the 
logarithmic  correction
exponent $r^{\rm{step}}$ to the Yang--Lee edge is shown as a 
function of  the assumed
critical coupling $\beta_c^{\rm{step}}$ and 
(b) the corresponding $\chi^2/{\rm{dof}}$.}
\label{fig09:step:r}
\end{figure} 

\begin{figure}[htb]
\vspace{13cm}
\includegraphics{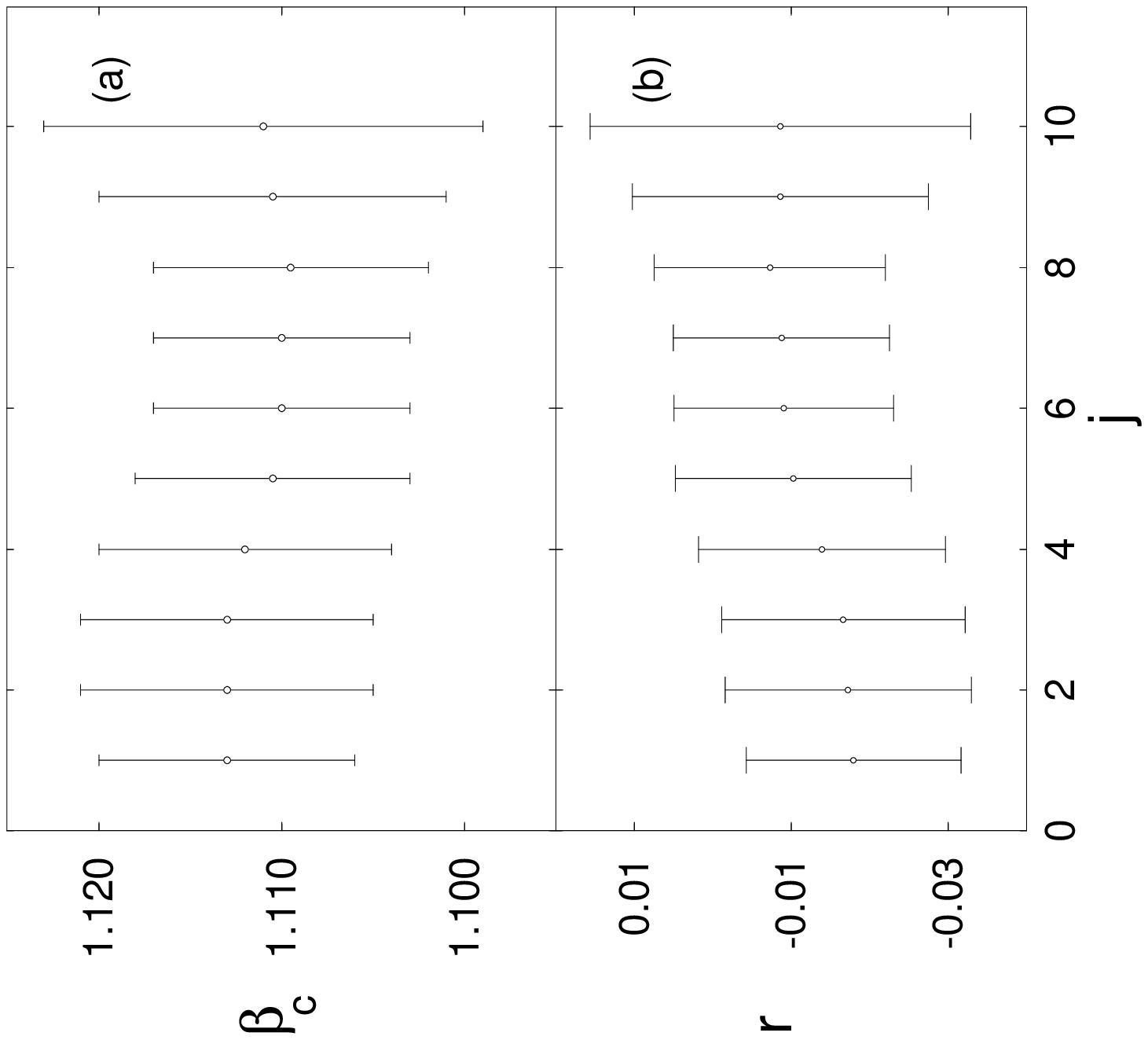}
\caption{Minimum and maximum acceptable (a) $\beta$ 
and (b) correction exponent $r$
from the $r$--analysis with criterion (\ref{criterion})
for the $XY$--model.}
\label{fig10:xy:j}
\end{figure}

\begin{figure}[htb]
\vspace{13cm}
\includegraphics{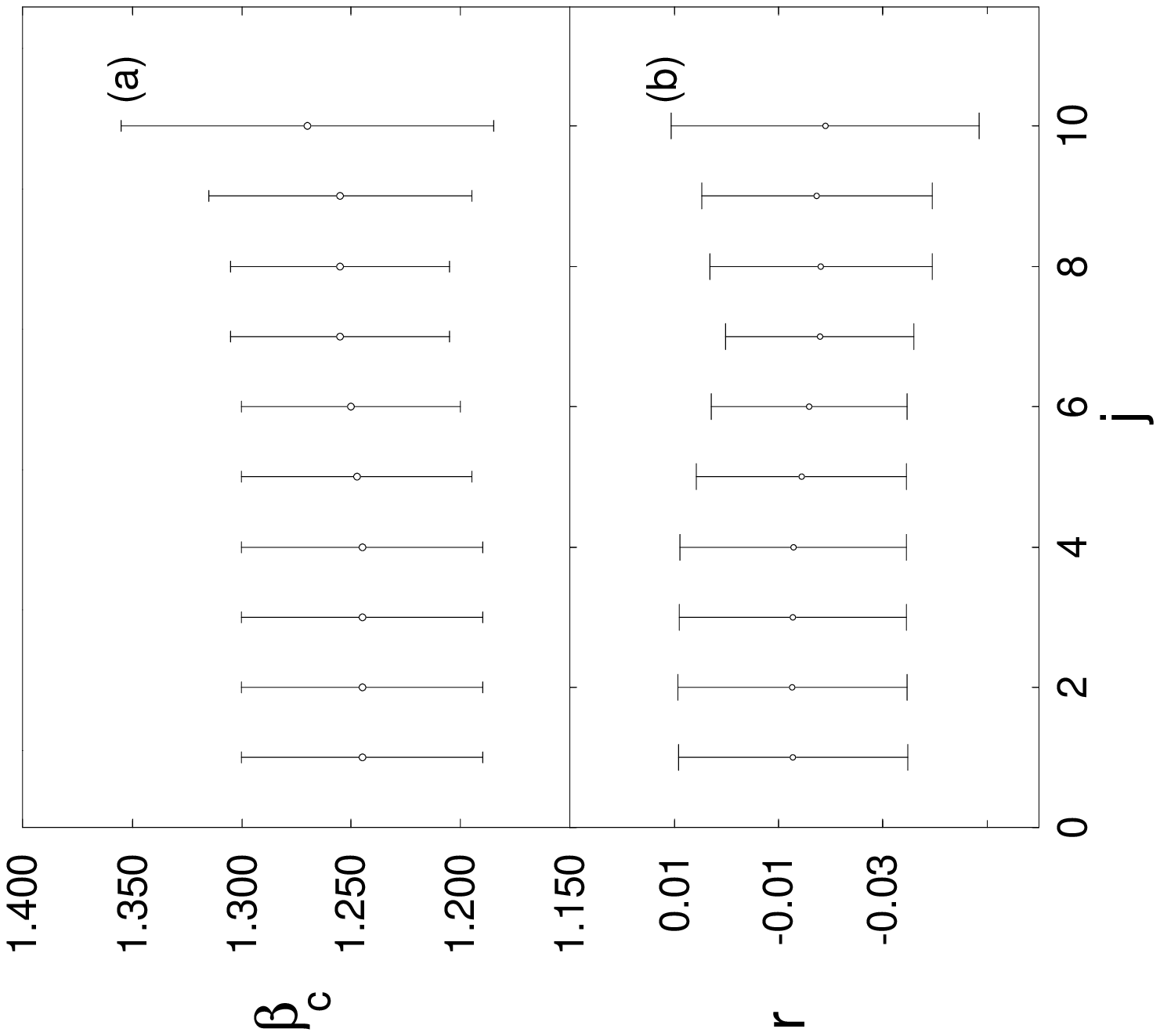}
\caption{Minimum and maximum acceptable (a) $\beta$
and (b) correction exponent $r$ 
from the $r$--analysis with criterion (\ref{criterion})
for the step model.}
\label{fig12:step:j}
\end{figure}

\begin{figure}[htb]
\vspace{13cm}       
\includegraphics{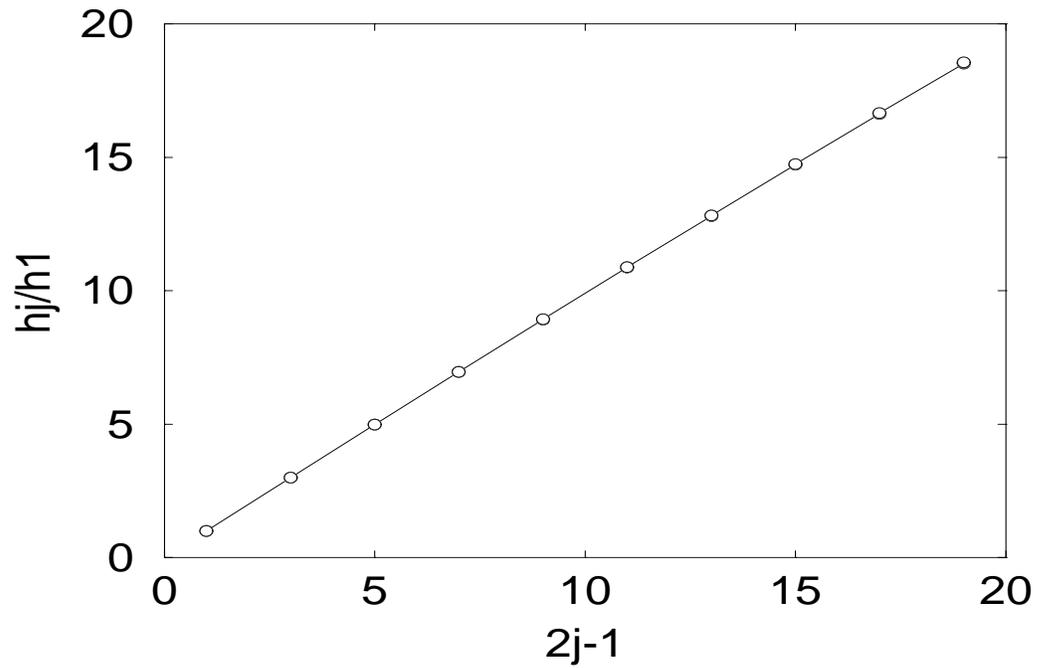}
\caption{Index scaling in the $XY$--model at $\beta_c = 1.115$.
The errors are smaller than the symbols.
Each symbol is in fact an overlap of four data points coming
from the four lattice sizes demonstrating excellent $L$--collapse.}
\label{fig11:xy:index}
\end{figure}

\end{document}